\documentclass[letterpaper,12pt]{article}

\usepackage{endnotes}
\usepackage{amssymb}
\usepackage{amsfonts}
\usepackage{amsmath}
\usepackage{amsthm}
\usepackage{graphicx,color} 

\usepackage{comment}
\usepackage{chicago}
\usepackage{array}

\usepackage{url}

\usepackage{setspace}
\def\R{\mathbb{R}}
\def\endproof{\hfill\diamondsuit}

\def\tv{\tilde{v}}
\def\ta{\tilde{a}}

\def\E{\mathbb{E}}
\def\V{\mathbb{V}}

\def\N{\mathbb N}

\newcommand{\hp}{\hat{p}}
\newcommand{\hy}{\hat{y}}
\newcommand{\hq}{\hat{q}}
\newcommand{\hz}{\hat{z}}

\numberwithin{equation}{section}
\theoremstyle{plain}                
\newtheorem{theorem}{Theorem}[section]
\newtheorem{lemma}[theorem]{Lemma}

\theoremstyle{definition}           
\newtheorem{definition}[theorem]{Definition}

\theoremstyle{remark}               

\usepackage[margin=1.25in]{geometry}

\usepackage{ulem} 

\begin{document}
\pagenumbering{arabic}
\pagestyle{empty}

\begin{center}
\large{\bf Information and Trading Targets in a Dynamic Market Equilibrium}\footnote{The authors benefited from helpful comments from Steve Shreve, Mihai Sirbu, Matthew Spiegel, Gordan {\v Z}itkovi\'c, and seminar participants at the 2015 NYU Microstructure Conference. The second  author has been supported by the National Science Foundation under Grant No. DMS-1411809 (2014 - 2017). Any opinions, findings,  and conclusions or recommendations expressed in this material are those of the authors and do not necessarily reflect the views of the National Science Foundation (NSF). Corresponding author: Duane Seppi, Tepper School of Business, Carnegie Mellon University, Pittsburgh, PA 15213. Phone: 412-268-2298. Email: ds64@andrew.cmu.edu.
}
\end{center}
\vspace{0.25cm}
\begin{center}

Jin Hyuk Choi\\ University of Texas at Austin

\ \\

Kasper Larsen \\  Carnegie Mellon University

\ \\

Duane J. Seppi\\ Carnegie Mellon University

\end{center}

\begin{center}
\ \\
 \today

\ \\

\end{center}

\begin{verse}
{\sc Abstract}:
This paper investigates the equilibrium interactions between trading targets and private information in a multi-period Kyle (1985) market.  There are two heterogenous investors who each follow dynamic trading strategies: A strategic portfolio rebalancer engages in order splitting to reach a cumulative trading target, and an unconstrained strategic insider trades on long-lived information.  We consider a baseline case in which the  rebalancer is initially  uninformed and also cases in which the rebalancer is initially partially informed.  We characterize a linear Bayesian Nash equilibrium, describe an algorithm for computing such equilibria, and present numerical results on properties of these equilibria.  
\end{verse}
\begin{verse}
\end{verse}

\newpage
\pagestyle{plain}
\addtocounter{page}{-1}

 Price discovery and liquidity in financial markets arise from the interactions of different investors with different information and trading motives using a variety of order execution strategies.\footnote{  The heterogeneity of the investing public is an important fact underlying current debates about high frequency trading (SEC  2010).} An important insight from Akerlof (1970), Grossman and Stiglitz (1980), Kyle (1985), and Glosten and Milgrom (1985) is that trading noise plays a critical role in markets subject to adverse selection when some investors trade on superior private information.  However, orders from investors with non-informational reasons to trade --- index funds, passive pension and insurance portfolios --- also presumably reflect optimizing behavior such as minimizing trading costs, optimizing hedging, and other portfolio structuring objectives. In addition, large passive investors routinely use optimized order execution algorithms to trade dynamically in current markets (see, e.g., Johnson 2010).

Our paper is the first to model a market equilibrium with dynamic trading by both informed and rebalancing investors without exogenous restrictions on information life. We specifically investigate a multi-period Kyle (1985) market in which there are two strategic investors with different trading motives who each follow optimal but different dynamic trading strategies.  One investor is a standard Kyle strategic informed investor with long-lived information. The other investor is a strategic portfolio rebalancer who trades over multiple rounds to minimize the cost of hitting a terminal trading target.  In addition, the model has noise traders and competitive market makers.  In our model, the informed investor's orders are masked by two types of trading noise over time: Independently and identically distributed noise trader orders and autocorrelated randomness in the  rebalancer's optimally chosen orders.

Our main results are:

\begin{itemize}
\item Sufficient conditions for a linear Bayesian Nash equilibrium are characterized.
\item An algorithm for computing such equilibria numerically is provided.
\item The presence of the rebalancer introduces several new features: i) the aggregate order flow is autocorrelated,  ii) expected trading volume for the insider and rebalancer is $U$-shaped over time, and iii) the price impact of the order flow is $S$-shaped with initial price impacts above those in Kyle and later price impacts below Kyle's.


\item The rebalancer's and insider's orders tend to become negatively correlated over time. As a result, their orders partially offset each other so that, on average, they provide liquidity to each other symbiotically with a reduced price impact.
\end{itemize}

Our analysis integrates two literatures on pricing and trading. The first literature is research on price discovery.  Kyle (1985) described equilibrium pricing and dynamic trading in a market with noise traders and a single investor who has long-lived private information.  Subsequent work by Holden and Subrahmanyam (1992), Foster and Viswanathan (1996), and Back, Cao, and Willard (2000) extended the model to allow for multiple informed investors with long-lived information.

A second literature studies optimal dynamic order execution for uninformed investors with trading targets.  This work includes Bertsimas and Lo (1998), Almgren and Chriss (1999, 2000), Gatheral and Scheid (2011),  Engel, Ferstenberg, and Russell (2012) and Predoiu, Shaikhet, and Shreve (2011) as well as  Bunnermeier and Pedersen (2005) on  predatory trading in response to predictable uninformed trading. This research all takes the price impact function for orders as an exogenously specified model input.   In contrast, we model optimal order execution in an equilibrium setting that endogenizes the price impact of orders and that reflects, in particular, the impact of strategic uninformed trading on price impacts.\footnote{In our model, order flow has a price impact due to adverse selection because of the insider's private information. Alternatively, one could model price impacts due to inventory costs and imperfect competition in liquidity provision.} Keim and Madhavan (1995) give empirical evidence on dynamic order-splitting by institutional investors.

Models combining both informed trading and optimized uninformed rebalancing have largely been restricted to static settings or to multi-period settings with short-lived information and/or exogenous restrictions on the rebalancer's trading strategies.  Admati and Pfleider (1988) study a dynamic market consisting of a series of repeating one-period trading rounds with short-lived information and uninformed discretionary liquidity traders who only trade once but decide when to time their trading.  An exception is Seppi (1990) who models an informed investor and a strategic uninformed investor with a trading target in a market in which both can trade dynamically.  His model is solved for separating and partial pooling equilibria with upstairs block trading, but only for a restricted set of particular model parameterizations.

Our paper is related to Degryse, de Jong, and van Kervel  (DJK 2014). Both papers model dynamic order splitting by an uninformed investor in a multi-period market.  Consequently, both models have autocorrelated (predictable) order flows because of the dynamic rebalancing. Order flow autocorrelation is empirically significant but absent in previous Kyle models.\footnote{For empirical evidence on order flow autocorrelation, see Hasbrouck (1991a,b) and also related empirical references in Degryse, de Jong, and van Kervel (2014).} However, there are two differences between our model and DJK (2014). First, the informed investors in DJK (2014) have short-lived private information --- i.e., they only have one chance to trade on high-frequency value innovations before they become public --- whereas our insider trades on long-lived information over multiple intra-day time periods.  Consequently, it is harder in our model to distinguish sequences of informed orders from sequences of uninformed orders. Second, our rebalancer's orders depend dynamically on the realized path of aggregate orders as well as on his rebalancing target, whereas the DJK (2014) rebalancer trades deterministically over time given his target. In particular, our rebalancer learns about the insider's information, since he can filter aggregate order flow better than the market makers. He then exploits this information in his trading. Our analysis is possible because we use the approach of Foster and Vishwanathan (1996) to circumvent the large state space problem mentioned in DJK (2014).

"Sunshine trading" is a prominent feature of models on uninformed rebalancing.  One version of sunshine trading exploits dynamic fluctuation in the price impacts of orders as the supply of liquidity is temporarily depleted and then replenished over time (see Predoiu, Shaikhet, and Shreve 2011). Another version  involves predictability in the timing of uninformed trading (see Admati and Pfleiderer 1988). Yet another version, new in DJK (2014) and our model, is that predictable order flow has no incremental information content and thus, absent frictions in the supply of liquidity, has no price impact. Thus, the rebalancer can use early trading to signal later trading. However, the numerical importance of sunshine trading is not large in our model.  This is because, unlike in DJK (2014), our insider trades dynamically. Other interactions with the insider, however, can reduce the rebalancer's trading costs at various times. These interactions include dynamic learning effects and also symbiotic liquidity provision when the insider's and rebalancer's orders are negatively correlated. The symbiotic liquidity provision is a new theoretical property of our model. Simply looking at the rebalancer's problem in a partial equilibrium analysis might cause one to miss these equilibrium considerations.



%

\section{Model}
We model a multi-period discrete-time market for a risky stock. A trading day is normalized to the interval $[0,1]$ during which there are $N\in\N$ time points at which trade can occur where $\Delta:= \frac1N>0$ is the time step. As in Kyle (1985), the stock's true value $\tv$ becomes publicly known at time $N+1$ after the market closes  at the end of the day. The value $\tv$ is normally distributed with mean zero and variance $\sigma^2_{\tv}>0$. Additionally, there is a money market account that pays a zero interest rate.

Four types of investors trade in our model:

\begin{enumerate}

\item An informed trader (who we call the insider) knows the true stock value $\tv$ at the beginning of trading and has  zero initial positions in both the stock and the money market account. The insider is risk-neutral and maximizes the expected value of her final wealth. The insider's order for the stock at time $n$, $n=1,...,N$, is denoted by $\Delta \theta^I_n$ where $\theta^I_n$ is her accumulated total stock position at time $n$.

\item A constrained investor (who we call the rebalancer) needs to rebalance his portfolio by buying or selling stock to reach a terminal trading target constraint $\ta$ on his ending stock position $\theta^R_N$ by the end of the trading day. For example, he might be the portfolio manager for a large index fund or a passive pension plan or an insurance company who needs to rebalance his portfolio.  In practice, such investors trade dynamically using optimal order execution algorithms to minimize their rebalancing costs. He starts the day with zero initial positions in both the stock and the money market account.\footnote{ Both the insider and the rebalancer finance their stock trading by borrowing/lending.   This assumption simplifies the notation for their objective functions but is without loss of generality.} The target $\ta$ is jointly normally distributed with the stock value $\tv$, has a mean of zero, a variance $\sigma^2_{\ta}>0$, and a correlation $\rho\in[0,1]$ with $\tv$. When $\rho$ is 0, the rebalancer is initially uninformed.  However, if $\rho > 0$, then we can think of the rebalancer as being initially informed about $\tv$ but subject to random binding non-public risk limits.\footnote{The fact that the terminal value $\tv$ is measured in dollars while the trading target $\ta$ is measured in shares is not problematic for $\tv$ and $\ta$ being correlated random variables.} The rebalancer is risk-neutral and maximizes the expected value of his final wealth subject to the terminal stock position constraint. The rebalancer's order for the stock at time $n$, $n=1,...,N$, is denoted by $\Delta \theta^R_n$, and the terminal constraint requires $\Delta \theta^R_N = \ta-\theta_{N-1}^R$ at time $N$.

\item  Noise traders submit net stock orders at times $n$, $n=1,...,N$, that are exogenously given by Brownian motion increments $\Delta w_n$. These increments are normally distributed with zero-mean and variance $\sigma_w^2 \Delta$ for a constant $\sigma_w>0$. We assume that $w$ is independent of $\tilde{v}$ and $\ta$.

\item Competitive risk-neutral market makers observe the aggregate net order flow $y_n$ at times $n$, $n=1,...,N$, where
\begin{align}\label{def:y_n}
y_n := \Delta \theta^I_n+\Delta \theta^R_n+\Delta w_n,\quad y_0 :=0.
\end{align}
Given competition and risk-neutrality, market makers clear the market (i.e., trade $-y_n$) at the stock price $p_n$ set to  be
\begin{align}\label{def:p_n}
p_n = \E[\tv|\sigma(y_1,...,y_n)],\quad n=1,2,...,N,\quad p_0=0,
\end{align}
where $\sigma(y_1,...,y_n)$ is the sigma-algebra generated by the order flow history.
\end{enumerate}

The constrained rebalancer's presence is the main difference between our setting and Kyle (1985) as well as the multi-agent settings in Holden and Subrahmanyam (1992) and Foster and Viswanathan (1996). As we shall see, the rebalancer's presence produces new stylized features such as autocorrelated order flow.

Because all initial positions are assumed to be zero (i.e., $\theta^I_0=\theta^R_0=0$), the insider chooses orders  $\Delta\theta_n^I \in \sigma(\tv,y_1,...,y_{n-1})$ at times $n$, $n=1,2,...,N,$ to maximize
\begin{align}\label{Insider_initial_problem}
\E\left[\theta^I_N(\tilde{v} -p_N) + \theta^I_{N-1}\Delta p_N +...+ \theta^I_{1}\Delta p_2\Big|\sigma(\tv)\right]=\E\left[\sum_{n=1}^N (\tilde{v} - p_n) \Delta \theta^I_n\Big|\sigma(\tv)\right].
\end{align}
On the other hand, the rebalancer faces the terminal constraint $\theta^R_N = \ta$. Therefore, he submits orders $ \Delta\theta_n^R \in \sigma(\ta,y_1,...,y_{n-1})$ at times $n$, $n=1,2,...,N-1$, to maximize
\begin{align}\label{Liquidator_initial_problem}
\E\left[\ta(\tilde{v} -p_N)+ \theta^R_{N-1}\Delta p_N +...+ \theta^R_{1}\Delta p_2\Big|\sigma(\ta)\right]=\frac{\rho\sigma_{\tv}}{\sigma_{\ta}}\ta^2-\E\left[\sum_{n=1}^N (\ta - \theta^R_{n-1}) \Delta p_n\Big|\sigma(\ta)\right],
\end{align}
given the trading target constraint $\theta_N^R = \ta$. The equality in \eqref{Liquidator_initial_problem} follows from $p_N= \sum_{n=1}^N \Delta p_n, \;p_0 = 0$, and $\E[\tv|\sigma(\ta)]=\frac{\rho\sigma_{\tv}}{\sigma_{\ta}}\ta$. As proven in the appendix, the insider's problem \eqref{Insider_initial_problem} and the rebalancer's problem   \eqref{Liquidator_initial_problem} are both quadratic optimization problems. We also note that the insider's, market makers', and rebalancer's information sets are not nested.

\begin{definition}\label{def:BayNash} A \emph{Baysian Nash} equilibrium is a collection of functions $(\theta^I_n, \theta^R_n, p_n)$  such that:
\begin{itemize}
\item[(i)] given the functions $(\theta^R_n, p_n)$, the strategy $\theta_n^I$ maximizes the insider's objective \eqref{Insider_initial_problem},
\item[(ii)] given the functions $(\theta^I_n, p_n)$, the strategy $\theta^R_n$ maximizes the rebalancer's objective \eqref{Liquidator_initial_problem},
\item[(iii)] given the functions $(\theta^I_n, \theta^R_n)$, the pricing rule $p_n$ satisfies \eqref{def:p_n}.
\end{itemize}
\end{definition}
To clarify this definition, we recall the Doob-Dynkin lemma: For any random variable $B$ and any $\sigma(B)$-measurable random variable $A$ we can find a deterministic function $f$ such that $A=f(B)$. Therefore, we can write $\theta^R_n = f^R_n(\ta,y_1,\ldots ,y_{n-1})$, $\theta^I_n = f^I_n(\tv, y_1,\ldots,y_{n-1})$, and $p_n = f^p_n(y_1, \ldots, y_n)$ for three deterministic functions $f^R_n$, $f^I_n$, and $f^p_n$. Definition \ref{def:BayNash} then means that the functions $f^R_n$, $f^I_n$, and $f^p_n$ are fixed whereas the realization of the aggregate order flow variables $y_1,...,y_n$ vary with the controls $\theta^I$ and $\theta^R$.

In what follows, our goal is to construct a linear Bayesian Nash equilibrium in which the following three properties hold: First, the insider's and rebalancer's optimal trading strategies take the forms:\footnote{If an additional term $\alpha^I_{n} q_{n-1}$ is included in the insider's strategy in \eqref{theta_I}, we find that $\alpha^I_n$ is zero in equilibrium.  Contact the authors for a proof of this result. }
\begin{align}
\Delta \theta^R_n &= \beta^R_n\Big(\ta-\theta^R_{n-1}\Big) + \alpha^R_{n} q_{n-1},\quad \theta^R_0=0,
\label{theta_L}\\
\Delta \theta^I_n &= \beta^I_n\Big(\tv-p_{n-1}\Big)
,\quad \theta^I_0=0, \label{theta_I}
\end{align}
 where $(\beta_n^R, \beta^I_n, \alpha^R_n)_{n=1}^N$ are constants with $\beta^R_N=1$ and $\alpha^R_N=0$. Second, the pricing rule has the dynamics
\begin{equation}\label{eq_p}
\Delta p_n = \lambda_n y_n + \mu_n q_{n-1},\quad p_0:=0,
\end{equation}
where $(\lambda_n, \mu_n)_{n=1}^N$ are constants. Third, the process $q_n$ has the dynamics
\begin{align}\label{def:q_n}
\Delta q_n = r_n y_n + s_nq_{n-1},\quad q_0:=0,
\end{align}
for constants $(r_n,s_n)_{n=1}^N$.  The rebalancer and insider are not restricted to use linear strategies like (\ref{theta_L}) and (\ref{theta_I}).  However, we will prove that they optimally choose such strategies in the equilibrium we construct.

The rebalancer's trading target $\ta$ necessitates the introduction of the process $q_n$ which is our model's main new feature. Much like $p_n$ is a state variable giving the market maker beliefs about the stock valuation, $q_n$ is a state variable indicating market maker beliefs about the rebalancer's remaining trading given the prior trading history. There are two things to note about $q_n$.  First, the rebalancer's trades are not limited to be a deterministic function of his target $\ta$. Rather, his trades can also depend on the realized prior order flow history as reflected in $q_n$.  This is in contrast to the  deterministic rebalancer trades in Degryse, de Jong, and van Kervel (2014). Second, if equations (\ref{theta_L}) through (\ref{def:q_n}) define a linear Bayesian Nash equilibrium, then the same equilibrium (with the same prices and orders) is obtained if $r_n$ and  $s_n$ are replaced with $x r_n$ and $x s_n$ and $\mu_n$ and $\alpha^R_{n}$ are replaced with $\mu_n/x$, and $\alpha^R_{n}/x$ 
for any scaler $x > 0$.  Thus, in the equilibrium considered below, we normalize $r_n$ and $s_n$ so that $q_n$ is the market  makers' expectation of the rebalancer's  remaining demand $\ta-\theta^R_n$ at time $n$ given the observed history of aggregate orders:\footnote{ An alternative scaling would be to set $q_{n}$ equal to $\E[y_n |\sigma(y_1,...,y_{n})]$.}
\begin{align}\label{hat_qn}
q_n= \E[\ta-\theta^R_n|\sigma(y_1,...,y_n)],\quad n=1,...,N.
\end{align}

The term $\ta-\theta^R_{n-1}$ in \eqref{theta_L} plays two roles in the rebalancer's strategy:  It is the distance between the rebalancer's current position and his final trading target $\ta$, and, in equilibrium, it is also private information that is useful in learning about possible stock price misvaluation $\tv - p_{n-1}$:
\begin{align}\label{eq:rebalancer-information-from-trading}
\begin{split}
\E[\tv -p_{n-1}|\sigma(\ta,y_1,...,y_{n-1})] &=\E[\tv -p_{n-1}|\sigma(\ta - \theta^R_{n-1}-q_{n-1},y_1,...,y_{n-1})] \\
&=\E[\tv -p_{n-1}|\sigma(\ta - \theta^R_{n-1}-q_{n-1})].
\end{split}
\end{align}
The first equality follows from $q_{n-1},\theta^R_{n-1}\in\sigma(\ta,y_1,...,y_{n-1})$. The second equality follows from the independence between $\tv -p_{n-1}$ and $y_1,...,y_{n-1}$ as well as the independence between $\ta - \theta^R_{n-1}-q_{n-1}$ and $y_1,...,y_{n-1}$. Thus, $\ta - \theta^R_{n-1}-q_{n-1}$ is, in general, incrementally informative about $\tv$ beyond the past order flow information already reflected in $p_{n-1}$. In particular, it is informative at $n>1$ even if $\rho=0$  (i.e., $\ta$ becomes informative about $\tv$ even if $\ta$ and $\tv$ are ex ante independent) because the rebalancer can filter the past order flow history to learn about $\tv$ better than the market makers.  This is a significant difference from deterministic rebalancing rules.

Similarly, the term $\tv-p_{n-1}$ in \eqref{theta_I} plays two roles in the insider's strategy:  It is  both private information about the stock value and, in equilibrium, informative about the remaining demand $\ta - \theta_{n-1}^R$ for the rebalancer:
\begin{align}\label{eq:insider-information}
\begin{split}
\E[&\ta -\theta^R_{n-1}|\sigma(\tv,y_1,...,y_{n-1})] \\&=q_{n-1}+\E[\ta -\theta^R_{n-1}-q_{n-1}|\sigma(\tv - p_{n-1},y_1,...,y_{n-1})] \\
&=q_{n-1}+\E[\ta -\theta^R_{n-1}-q_{n-1}|\sigma(\tv - p_{n-1})].
\end{split}
\end{align}
The first equality follows from $q_{n-1}, p_{n-1}\in \sigma(y_1,...,y_{n-1})$. The second equality follows from the independence between $\tv -p_{n-1}$ and $y_1,...,y_{n-1}$ as well as the independence between $\ta - \theta^R_{n-1}-q_{n-1}$ and $y_1,...,y_{n-1}$.

\subsection{Equilibrium}
In this section we characterize sufficient conditions for existence of a linear Bayesian Nash equilibrium of the form in \eqref{theta_L} through \eqref{def:q_n}.  The analysis follows the logic of Foster and Viswanathan (1996) closely.

To begin, we consider a complete set of possible candidate values for the equilibrium constants
\begin{align}\label{all_constants}
\begin{split}
&\lambda_n, \mu_n, r_n, s_n, \beta_n^R, \alpha_n^R, \beta_n^I, 
\quad n=1,\ldots,N,
\end{split}
\end{align}
with
\begin{align}\label{all_constantsN}
&\beta^R_N=1,\quad\alpha^R_N=0.
\end{align}
The restrictions in (\ref{all_constantsN}) at date $N$ reflect the fact that the rebalancer must achieve his target $\ta$ after his last round of trade. Our goal of this section is to identify sufficient conditions for a candidate set of specific coefficient values to be an equilibrium.  We do this in three steps.

The first step takes a set of candidate constants \eqref{all_constants}-\eqref{all_constantsN} and  computes (using the terminology and notation of Foster and Viswanathan 1996) the following system of ``hat'' price and order flow processes
\begin{align}
\Delta\hat{\theta}_n^I &:= \beta_n^I (\tv-\hp_{n-1}) 
\quad \hat{\theta}^I_0:=0,\label{optimal_nheta_I}\\
\Delta\hat{\theta}_n^R &:= \beta_n^R (\ta-\hat{\theta}^R_{n-1}) + \alpha_n^R \hq_{n-1},\quad \hat{\theta}^R_0:=0,\label{optimal_nheta_L}\\
\hy_n&:=\Delta\hat{\theta}_n^I+\Delta\hat{\theta}_n^R+\Delta w_n,\quad \hy_0:=0,\label{haty}\\
\Delta \hp_n& := \lambda_n \hy_n + \mu_n \hq_{n-1},\quad \hp_0 :=0,\label{hatp}\\
\Delta \hq_n &:=r_n \hy_n + s_n \hq_{n-1},\quad \hq_0 :=0\label{hatq}.
\end{align}
The system of processes $(\Delta\hat{\theta}_n^I,\Delta\hat{\theta}^R_n, \hy_n, \Delta \hp_n,\Delta \hq_n, )$ is fully specified (autonomous) by the coefficients \eqref{all_constants}. Furthermore, given the zero-mean and joint normality of $\tv$, $\ta$, and $w$, the ``hat'' system \eqref{optimal_nheta_I}-\eqref{hatq} is also zero-mean and jointly normal. We define the variances and covariance for the ``hat'' dynamics, $n= 0,1,...,N$, by\footnote{We note that $\Sigma^{(2)}_n$ must be non-increasing over time (as in Kyle 1985) but $\Sigma^{(1)}_n$ might not be.}
\begin{align}
\Sigma_{n}^{(1)}& := \V\big[\ta - \hat{\theta}_{n}^R - \hq_{n}\big]\label{S11},\\
\Sigma_{n}^{(2)}&:= \V[\tv-\hp_{n}\big]\label{S22},\\
\Sigma_{n}^{(3)}&:= \E \big[\big(\ta - \hat{\theta}_{n}^R -\hq_{n}\big)(\tv - \hp_{n})\big].\label{S33}
\end{align}
These moments are ``post-trade" at time $n$ in that they reflect the trading up-through and including the time $n$ order flow $y_n$. In other words, they are inputs for trading round $n+1$. The initial variances and covariance at $n = 0$ are exogenuously given by
\begin{align}\label{eq:starting-values}
\Sigma_0^{(1)}=\sigma^2_{\ta},\quad \Sigma_0^{(2)} =\sigma^2_{\tv}, \quad \Sigma_0^{(3)}=\rho.
\end{align}

The ``hat'' processes \eqref{optimal_nheta_I}-\eqref{hatq} will be used to make the problems \eqref{Insider_initial_problem} and \eqref{Liquidator_initial_problem} analytically tractable in the sense that both the insider's problem and the rebalancer's problem can be described by low dimensional state processes (see \ref{X} and \ref{Y} below). In particular, the ``hat'' processes denote the processes that agents believe other agents believe describe the equilibrium. In equilibrium, these beliefs must be correct.  This consistency requirement imposes two groups of conditions that a set of candidate constants \eqref{all_constants} must satisfy to be equilibrium constants. The next two steps explain these conditions.

The second step requires the coefficients $(\lambda_n,\mu_n,s_n,r_n)_{n=1}^N$ of the price and order flow state variable processes $(p_n,q_n)_{n=1}^N$ to be consistent in equilibrium with Bayesian updating.  In particular, if market makers believe that the insider and rebalancer are following the ``hat'' strategies, then we can re-write \eqref{def:p_n} as
\begin{align}\label{eq_pp}
\begin{split}
\Delta \hp_n
 &= \lambda_n \big(\hy_n - \E[\hy_n |\sigma(\hy_1,...,\hy_{n-1})] \big) \\
 &= \lambda_n \big(\hy_n - [\beta_n^R \, \E[\ta-\hat{\theta}^R_{n-1} |\sigma(\hy_1,...,\hy_{n-1})] + \alpha_n^R \hq_{n-1} ]\big) \\
 &= \lambda_n \big(\hy_n - (\alpha_n^R+\beta^R_n)\hq_{n-1} \big),
 \end{split}
\end{align}
for $n=1,...,N$. The first equality follows from the fact that, given the jointly Gaussian structure of the ``hat'' processes, conditional expectations are linear projections. The second equality follows from (i) the definition of the aggregate order flow \eqref{haty}, (ii) the independence between $\tv-\hp_{n-1}$ and past order flows, and (iii) the assumption that the noise trader orders are zero--mean and i.i.d. over time. The final equality follows from 
the normalization that $\hq_{n-1} = \E[\ta-\hat{\theta}^R_{n-1} |\sigma(\hy_1,...,\hy_{n-1})]$. Comparing the last line of \eqref{eq_pp} with \eqref{eq_p} and using that $\lambda_n$ equals the projection coefficient
\begin{align}\label{eq:projection-coef}
\frac{{\rm Cov}(\tv - \hp_{n-1}, \hy_n - \E[\hy_n |\sigma\hy_1,...,\hy_{n-1})])}{\V(y_n - \E[\hy_n |\sigma(\hy_1,...,\hy_{n-1})])}
\end{align}
gives restrictions on the coefficients of the price process in terms of the insider and rebalancer strategy coefficients.  A similar logic can also be used to derive restrictions on the coefficients of the $\hq_n$  process in terms of the investor strategy coefficients. These calculations lead to the following four restrictions on the state variable and strategy constants in a linear Bayesian Nash equilibrium for $n=1,...,N$ (see the proof of Lemma \ref{lem:Kalman} in Appendix \ref{sec:Kalman}):
\begin{align}
\lambda_n&=\frac{\beta_n^I \Sigma_{n-1}^{(2)} + \beta_n^R \Sigma_{n-1}^{(3)}}{  (\beta_n^I)^2 \Sigma_{n-1}^{(2)}+ (\beta_n^R)^2 \Sigma_{n-1}^{(1)} + 2 \beta_n^I \beta_n^R \Sigma_{n-1}^{(3)}  + \sigma_w^2\Delta},\label{ass1lambda}\\
r_n&=\frac{(1-\beta_n^R)\big(\beta_n^I \Sigma_{n-1}^{(3)} + \beta_n^R \Sigma_{n-1}^{(1)}\big)}{ (\beta_n^I)^2 \Sigma_{n-1}^{(2)}+ (\beta_n^R)^2 \Sigma_{n-1}^{(1)} + 2 \beta_n^I \beta_n^R \Sigma_{n-1}^{(3)}  + \sigma_w^2\Delta}, \label{ass1r}\\
\mu_n&=-\lambda_n(\alpha_n^R+\beta_n^R),\label{ass1mu}\\
s_n&=-(1+r_n)( \alpha_n^R+\beta_n^R).\label{ass1s}
\end{align}
Here the conditional variances and covariance from \eqref{S11}-\eqref{S33} can  recursively be computed as
\begin{align}
\Sigma^{(1)}_{n}  &=(1-\beta_n^R)\big( (1-\beta_n^R-r_n\beta_n^R)\Sigma_{n-1}^{(1)} - r_n \beta_n^I \Sigma_{n-1}^{(3)}  \big),\label{S1iterative}\\
\Sigma^{(2)}_{n} &=(1-\lambda_n \beta_n^I)\Sigma_{n-1}^{(2)} - \lambda_n \beta_n^R \Sigma_{n-1}^{(3)},\label{S2iterative}\\
\Sigma^{(3)}_{n} &=(1-\beta_n^R)\big(    (1-\lambda_n\beta_n^I) \Sigma_{n-1}^{(3)} - \lambda_n \beta_n^R \Sigma_{n-1}^{(1)}\big).\label{S3iterative}
\end{align}
We note the ``block'' structure: The values of the updating coefficients $\lambda_n$ and $r_n$ just depend on the strategy coefficients $\beta^R_n$ and $\beta^I_n$ at date $n$ and the incoming variances and covariance from time $n-1$ (along with the exogenous noise trading variance $\sigma_w^2$).  The post-trade variances and covariance $\Sigma^{(1)}_n$, $\Sigma^{(2)}_n$, and $\Sigma^{(3)}_n$ at time $n$ just depend on the updating coefficients $\lambda_n$ and $r_n$ at time $n$, the strategy coefficients at time $n$, and the prior variances and covariance from time $n-1$.  Lastly, $\mu_n$ and $s_n$ depend on $\lambda_n$ and $\mu_n$ as well as the rebalancer's set of  strategy coefficients $(\beta^R_n,\alpha^R_n)$.

The third and last step begins by deriving value functions for the two strategic investors (see \ref{Insider_initial_problem} and \ref{Liquidator_initial_problem}).  Consider first the insider at a generic time $n$. As in Foster and Viswanathan (1996), the insider not only knows the final stock value $\tv$, but also the extent to which the actual ``unhatted'' prices, quantity expectations, and rebalancer's positions (i.e., $p_n$, $q_n$, and $\theta^R_{n}$ given by \ref{theta_L}, \ref{eq_p}, and \ref{def:q_n}) given her actual trades $\Delta \theta^I_1, \ldots, \Delta \theta^I_{n}$ deviate from the ``hatted'' values $\hp_n$, $\hq_n$, and $\hat \theta^R_{n}$ given by \eqref{hatp}, \eqref{hatq}, and \eqref{optimal_nheta_L}) if she had instead traded according to the candidate ``hat'' insider process $\Delta \hat \theta^I_1, \ldots, \Delta \hat \theta^I_{n}$ in \eqref{optimal_nheta_I}.
In deriving the equilibrium, we need to allow for the possibility of past suboptimal play. Hence, the ``un-hatted'' variables are the variable values given her actual (potentially arbitrary) orders whereas the ``hat'' variables are not affected by actual orders. When the rebalancer's strategy is taken to be fixed by  \eqref{theta_L}, it is characterized by the two sequences of candidate coefficients $\beta^R_1, \ldots, \beta^R_N$ and $\alpha^R_1, \ldots, \alpha^R_N$. However, even though the rebalancer's strategy is fixed, its realizations are subject to the insider's choice of $\theta^I$ since the aggregate order flow affects the rebalancer's actual orders.  Similar statements apply for prices $p_n$ and $\hp_n$ and the quantity expectations $q_n$ and $\hq_n$.

Based on Foster and Viswanathan (1996), it would be natural to consider
\begin{align}\label{X_naive}
\hspace{-0.15in} \tv-\hp_n, \quad \hq_n, \quad \hat{\theta}_n^R - \theta_n^R, \quad \hq_n - q_n, \quad\hp_n - p_n.
\end{align}
as state variables for the insider's problem. However, we show that the following two composite state variables are sufficient statistics for the insider's value function:
\begin{align}\label{X}
\hspace{-0.1in}  X^{(1)}_{n} := \tv-p_n,\quad   X^{(2)}_{n} := (\hat{\theta}^R_{n} -\theta^R_{n}) + (\hq_{n}-q_{n}) +\tfrac{\Sigma^{(3)}_{n}}{\Sigma^{(2)}_{n}} \big(\tv- \hat{p}_{n}\big),\quad n=0,...,N.
\end{align}
From a technical point of view, this is a substantial reduction in the set of state variables from \eqref{X_naive}.  This surely seems like the minimum number of state variables necessary for the insider's problem. Lemma \ref{Iinfo} in Appendix \ref{sec:proofs} ensures that these processes are observable for the insider. From \eqref{X}, we see that in equilibrium, with $p_n = \hp_n$, $q_n = \hq_n$, and $\theta^R_{n} = \hat{\theta}^R_{n}$, we have the relation
\begin{align}\label{X1X2equilibrium}
 X^{(2)}_{n} = \tfrac{\Sigma^{(3)}_{n}}{\Sigma^{(2)}_{n}} X^{(1)}_{n},\quad n=0,1...,N.
\end{align}

Lemma \ref{Iinfo} in Appendix \ref{sec:proofs} shows that the insider's value function for $n=0,1,...,N$ has the quadratic form
\begin{equation}\begin{split}\label{Ivalue-body}
\max_{\stackrel{\Delta\theta_k^I\in \sigma(\tv,y_1,...,y_{k-1})}{n+1\leq k \leq N}} &\E \Big[ \sum_{k=n+1}^N (\tv-p_k)\Delta \theta^I_k \Big|\sigma( \tv,y_1,...,y_n )\Big]\\&= I_n^{(0)} +I_n^{(1,1)} (X_n^{(1)})^2+I_n^{(1,2)} X_n^{(1)}X_n^{(2)}+I_n^{(2,2)} (X_n^{(2)})^2,
\end{split}\end{equation}
where $I_n^{(0)},I_n^{(1,1)},I_n^{(1,2)}$, and $I_n^{(2,2)}$ are constants.
Furthermore, Lemma \ref{Iinfo} also shows that the insider's problem \eqref{Ivalue-body} is quadratic in $\Delta\theta^I_n$. The first-order-condition for \eqref{Ivalue-body} produces the  insider's optimal order candidate process
\begin{equation}\begin{split}\label{Iopt}
\Delta\theta_n^I =
\gamma^{(1)}_n X_{n-1}^{(1)}+\gamma^{(2)}_n X_{n-1}^{(2)}, \quad n=1,...,N,
\end{split}\end{equation}
where
\begin{align}
\gamma_n^{(1)}&:= \tfrac{-1 + I^{(1,2)}_n r_n +
 2 I^{(1,1)}_n\lambda_n}{2 (I^{(2,2)}_n r_n^2 + \lambda_n (-1 + I^{(1,2)}_n r_n +
  I^{(1,1)}_n\lambda_n))},
\label{Iopt1} \\
\gamma_n^{(2)}&:= -\beta^R_n + \tfrac{-2 I^{(2,2)}_n r_n (-1 + \beta^R_n) +
    I^{(1,2)}_n \lambda_n -\beta^R_n\lambda_n (I^{(1,2)}_n+1)
}{2 (I^{(2,2)}_n r_n^2 + \lambda_n (-1 + I^{(1,2)}_n r_n +
  I^{(1,1)}_n\lambda_n))}.
\label{Iopt2}
\end{align}
The associated second-order condition for the insider's optimal strategy is
\begin{align}
I^{(2,2)}_n r_n^2 + I^{(1,2)}_n r_n \lambda_n + I^{(1,1)}_n \lambda_n^2 < \lambda_n. \label{Isecond}
\end{align}
By inserting the insider's candidate strategy \eqref{Iopt}-\eqref{Iopt2} into the expectation in \eqref{Ivalue-body}, we can determine the insider's value function coefficients recursively. More specifically, the expectation is computed in equation \eqref{Iconditional1} Appendix \ref{sec:proofs} and the resulting recursions are given by \ref{I^{(1,1)}_n}-\ref{I^{(2,2)}_n} in Appendix \ref{sec:recursion}.

By equating the coefficients in (\ref{Iopt}) with (\ref{theta_I}) and using the equilibrium condition \eqref{X1X2equilibrium}, we get the following condition on the insider's strategy coefficient
\begin{align}\label{eq:betaI}
\beta^I_n = \gamma^{(1)}_n +  \gamma^{(2)}_n\tfrac{\Sigma^{(3)}_{n-1}}{\Sigma^{(2)}_{n-1}},\quad n=1...,N.
\end{align}
For fixed $\Sigma^{(1)}_{n},...,\Sigma^{(3)}_{n}$, we can use the linear equations \eqref{S1iterative}-\eqref{S3iterative} to express $\Sigma^{(1)}_{n-1},...,$ $\Sigma^{(3)}_{n-1}$ in terms of $r_n,\lambda_n,\beta_n^I,\beta^R_n$. Equations (\ref{Iopt1})-(\ref{Iopt2}) and (\ref{ass1lambda})-(\ref{ass1r}) can then be used to see that \eqref{eq:betaI} is a fifth--degree polynomial in $(\beta^R_n,\beta^I_n)$ whenever $\Sigma^{(i)}_{n}$, $i=1,2,3$, and $I^{(i,j)}_n$, $i=1,2$ and $i\le j\le 2$, are fixed.

We next turn to the rebalancer's problem.  Again, based on Foster and Viswanathan (1996) it would be natural to consider
\begin{align}\label{Y_naive}
\ta-\hat{\theta}_n^R, \quad \hq_n, \quad \hat{\theta}_n^R - \theta_n^R, \quad \hq_n - q_n, \quad \hp_n - p_n,
\end{align}
as the rebalancer's state variables. However, now just three composite state variable are sufficient statistics for the rebalancer's value function
\begin{align}\label{Y}
Y_n^{(1)}:=\ta-\theta_n^R, \quad   Y_n^{(2)}:= (\hp_n - p_n) + \tfrac{\Sigma^{(3)}_n}{\Sigma_n^{(1)}}(\ta - \hat{\theta}_n^R -\hq_n ),\quad  Y_n^{(3)}:=q_n,\quad n=0,1,...,N.
\end{align}
Lemma \ref{Linfo} in Appendix \ref{sec:proofs} ensures that these processes are observable for the rebalancer.  Based on \eqref{Y}, we see that in equilibrium, with  $p_n = \hp_n$, $q_n = \hq_n$, and $\theta^I_{n} = \hat{\theta}^I_{n}$, we have the relation
\begin{align}\label{Y1Y2Y3equilibrium}
 Y^{(2)}_{n} = \tfrac{\Sigma^{(3)}_{n}}{\Sigma^{(1)}_{n}} (Y^{(1)}_{n}-Y^{(3)}_{n}),\quad n=1,...,N.
\end{align}
When the insider's strategy is fixed as in \eqref{theta_I},  Lemma \ref{Linfo} in Appendix \ref{sec:proofs} shows that the rebalancer's value function becomes
\begin{equation}\begin{split}\label{Lvalue-body}
\max_{\stackrel{\Delta\theta_k^R\in \sigma(\ta,y_1,...,y_{k-1})}{n+1\leq k \leq N-1}} &-\E \Big[ \sum_{k=n+1}^N (\ta-\theta_{k-1}^R)\Delta p_k \Big|\sigma( \ta,y_1,...,y_n )\Big] \\&=  L_n^{(0)} +\sum_{1\le i\le j\le3}L_n^{(i,j)} Y_n^{(i)}Y_n^{(j)},
\end{split}
\end{equation}
where $L^{(0)}_n,...,L^{(3,3)}_n$ are constants.  Lemma \ref{Linfo} also ensures that the rebalancer's problem \eqref{Lvalue-body} is quadratic in $\Delta \theta_n^R$. The corresponding first-order-condition produces the candidate optimizer for the rebalancer's order
\begin{equation}\begin{split}\label{Lopt}
\Delta\theta_n^R =
 \delta_n^{(1)}Y_{n-1}^{(1)}+ \delta_n^{(2)}Y_{n-1}^{(2)}+ \delta_n^{(3)}Y_{n-1}^{(3)}, \quad n=1,...,N,
\end{split}\end{equation}
where
\begin{footnotesize}
\begin{align}
\delta_n^{(1)}&:= \frac{2 L^{(1,1)}_n - L^{(1,3)}_n r_n + \lambda_n +
 L^{(1,2)}_n\lambda_n}{2 \big(L^{(1,1)}_n - L^{(1,3)}_n r_n +
   L^{(3,3)}_n r_n^2 + \lambda_n (L^{(1,2)}_n - L^{(2,3)}_n r_n + L^{(2,2)}_n\lambda_n)\big)},
\label{Ropt1} \\
\delta_n^{(2)}&:= -\beta^I_n + \frac{
 L^{(1,2)}_n - r_n (L^{(2,3)}_n + L^{(1,3)}_n\beta^I_n) + L^{(1,2)}_n\beta_n^I \lambda_n + 2 (L^{(1,1)}_n \beta_n^I + L^{(2,2)}_n \lambda_n)}{2\big(L^{(1,1)}_n - L^{(1,3)}_n r_n +
   L^{(3,3)}_n r_n^2 + \lambda_n (L^{(1,2)}_n - L^{(2,3)}_n r_n + L^{(2,2)}_n\lambda_n)\big)},
\label{Ropt2} \\
\begin{split}
\delta_n^{(3)}&:= \frac{ \begin{array}{c}
\Big(-2 L^{(3,3)}_n r_n - L^{(1,3)}_n (-1 + \alpha_n^R + r_n\alpha^R_n + \beta^R_n + r_n\beta^R_n) +
 L^{(2,3)}_n \lambda_n \\
 + (\alpha^R_n + \beta_n^R) \big(2 L^{(3,3)}_n r_n (1+r_n) + \lambda_n (L^{(1,2)}_n - L^{(2,3)}_n - 2 L^{(2,3)}_n r_n + 2 L^{(2,2)}_n\lambda_n)\big)\Big)\end{array} }{
2 \big(L^{(1,1)}_n - L^{(1,3)}_n r_n +
   L^{(3,3)}_n r_n^2 + \lambda_n (L^{(1,2)}_n - L^{(2,3)}_n r_n + L^{(2,2)}_n\lambda_n)\big)}. \label{Iopt3}
\end{split}
\end{align}
\end{footnotesize}
The associated second-order condition for the rebalancer's optimal strategy is
\begin{align}
L^{(1,1)}_n + L^{(3,3)}_n r_n^2 + L^{(1,2)}_n \lambda_n +  L^{(2,2)}_n \lambda_n^2 < L^{(1,3)}_n r_n + L^{(2,3)}_n r_n\lambda_n.
\label{Lsecond}
\end{align}
Similar to the insider's problem, by inserting the rebalancer's candidate strategy \eqref{Lopt}-\eqref{Iopt3} into the expectation in \eqref{Lvalue-body} we can find the rebalancer's value function coefficients recursively (see equations \ref{L11}-\ref{L33} in Appendix \ref{sec:recursion}).

By equating the coefficients in (\ref{Lopt}) with (\ref{theta_L}) and using the equilibrium condition \eqref{Y1Y2Y3equilibrium} we get the requirements
\begin{align}\label{eq:betaRalpha}
\beta^R_n = \delta^{(1)}_n +  \delta^{(2)}_n\tfrac{\Sigma^{(3)}_{n-1}}{\Sigma^{(1)}_{n-1}},\quad \alpha^R_n = \delta^{(3)}_n -  \delta^{(2)}_n\tfrac{\Sigma^{(3)}_{n-1}}{\Sigma^{(1)}_{n-1}},\quad n=1,...,N.
\end{align}
Similarly to \eqref{eq:betaI}, the first part of \eqref{eq:betaRalpha} is fifth--degree polynomial in $(\beta^R_n,\beta^I_n)$ whenever $\Sigma^{(i)}_{n}$, $i=1,2,3$, and $L^{(i,j)}_n$, $i=1,2,3$ and $i\le j\le 3$, are fixed.

Our main theoretical result is the following:

\begin{theorem}\label{thm:main}  If the constants \eqref{all_constants} and the associated terms
\begin{align}\label{thm_constants}
\Sigma^{(1)}_{n},\Sigma^{(2)}_{n},\Sigma^{(3)}_{n},
( I_n^{(i,j)})_{1\leq i \leq j \leq 2},
(L_n^{(i,j)})_{1\leq i \leq j \leq 3},\quad n=1,...,N,
\end{align}
satisfy the pricing coefficient relations \eqref{ass1lambda}-\eqref{ass1s}, the variances and covariance recursions \eqref{S1iterative}-\eqref{S3iterative}, the rebalancer's target constraint \eqref{all_constantsN}, the value function coefficient recursions \eqref{I^{(1,1)}_n}-\eqref{I^{(2,2)}_n} and \eqref{L11}-\eqref{L33}, the second-order-conditions
\eqref{Isecond} and \eqref{Lsecond} as well as the equilibrium conditions \eqref{eq:betaI} and \eqref{eq:betaRalpha}, then a linear Bayesian Nash equilibrium exists of the form given in \eqref{theta_L}-\eqref{def:q_n}.
Furthermore, we have
\begin{align}\label{zeroIs}
&r_N = 0,\quad \mu_N = -\lambda_N,\quad s_N =-1,\quad
\beta_N^I =
\big(\frac{1}{2\lambda_N} - \frac{\Sigma_{N-1}^{(3)} }{2 \Sigma_{N-1}^{(2)}} \big),\quad \lambda_N>0.
\end{align}
\end{theorem}

This characterization result is the analogue of Proposition 1 in Foster and Viswa-nathan (1996).  As already noted, the new feature in our model, compared to Foster and Viswanathan (1996) and Kyle (1985), is the presence of the $q_n$ process in the equilibrium price dynamics \eqref{eq_p}. This produces new stylized features including autocorrelation of the equilibrium aggregate order flow:
\begin{align}\label{autocor}
\begin{split}
\E&[y_n|\sigma(y_1,...,y_{n-1})] \\
&= \E[\Delta\theta^I_n+\Delta\theta^R_n + \Delta w_n|\sigma(y_1,...,y_{n-1})]\\
&= \alpha^R_n \, q_{n-1}+\E[\beta_n^I(\tv -p_{n-1}) +\beta_n^R(\ta -\theta^R_{n-1}) |\sigma(y_1,...,y_{n-1})] \\
&=(\alpha^R_n+\beta^R_n) \, q_{n-1},
\end{split}
\end{align}
which, in general, is not zero. 
The last equality follows, in part, from the earlier observation that, in equilibrium, $q_{n-1}$ is the conditional expectation of $\ta -\theta^R_{n-1}$ given the prior trading history.

\subsection{Algorithm}\label{sec:algorithm}
This section describes an algorithm for searching numerically for a linear Bayesian Nash equilibrium. The algorithm is similar in logic to the algorithm in Section V in Foster and Viswanathan (1996), except that our algorithm requires three constants as inputs (due to the presence of two strategic agents) whereas Foster and Viswanathan (1996) only requires one constant as an input.

The algorithm starts by taking as inputs three conjectured conditional moments for the final time $N$ round of trading:\footnote{We do not take the post-trade date $N$ moments $(\Sigma_N^{(1)},\Sigma_N^{(2)},\Sigma_N^{(3)})$ as inputs because they are after the last round of trading.  In addition, \eqref{S1iterative} and \eqref{S3iterative} together with the terminal condition $\beta^R_N=1$ imply that $\Sigma_N^{(1)}=\Sigma_N^{(3)}=0$.}
\begin{align}\label{Sigma_N}
\Sigma_{N-1}^{(1)} >0,\quad \Sigma_{N-1}^{(2)} >0,\quad \Sigma_{N-1}^{(3)}\in\R\quad\text{such that}\quad\big(\Sigma_{N-1}^{(3)}\big)^2 \le\Sigma_{N-1}^{(1)}\Sigma_{N-1}^{(2)}.
\end{align}
The algorithm then proceeds through backward induction.

\bigskip \noindent{\bf Starting step for trading time $N$:} We need $(\lambda_N,\beta^I_N)$ to satisfy \eqref{ass1lambda} for $n=N$ and the last two parts of  \eqref{zeroIs}. Given those two constants $(\lambda_N,\beta^I_N)$, we can define
\begin{align}
&\beta^R_N:=1,\quad\alpha^R_N:= r_N:=0,\quad \mu_N := - \lambda_N,\quad s_N := -1.
\end{align}
Because of the rebalancer's terminal constraint, his last round of trading (i.e., at time $N$) does not involve any optimization, and so we have
\begin{align*}
\E\left[-(\ta -\theta^R_{N-1})\Delta p_N|\sigma(\ta,y_1,...,y_{N-1})\right]& = -Y_{N-1}^{(1)}\big( \lambda_N(Y^{(1)}_{N-1} + \beta^I_NY^{(2)}_{N-1}) -\lambda_N Y^{(3)}_{N-1}\big).
\end{align*}
This relation implies that the rebalancer's value function coefficients for $n=N-1$ are given by
\begin{align}
L^{(1,1)}_{N-1} = -\lambda_N,\, L^{(1,2)}_{N-1}=-\lambda_N\beta^I_{N},\, L^{(1,3)}_{N-1} = \lambda_N,\, L^{(2,2)}_{N-1}=L^{(2,3)}_{N-1}=L^{(3,3)}_{N-1}=0.
\end{align}
On the other hand, the insider's problem in the last round of trading is the similar to her problem in any other round of trading. By inserting the boundary conditions
$$
I^{(1,1)}_{N}=I^{(1,2)}_{N}=I^{(2,2)}_{N}=0
$$
into the recursions \eqref{I^{(1,1)}_n}-\eqref{I^{(2,2)}_n}, we produce the value function coefficients $
I^{(i,j)}_{N-1}
$.\ \\

\noindent{\bf Induction step:} At each time $n$ the algorithm takes the following terms as inputs:
\begin{align}
\Sigma^{(1)}_{n},\Sigma^{(2)}_{n},\Sigma^{(3)}_{n},
( I_n^{(i,j)})_{1\leq i \leq j \leq 2}, 
(L_n^{(i,j)})_{1\leq i \leq j \leq 3}.
\end{align}
We first find the constants $(\lambda_n,r_n,\Sigma^{(1)}_{n-1},\Sigma^{(2)}_{n-1},\Sigma^{(3)}_{n-1},\beta_n^I,\beta^R_n)$ by requiring that \eqref{ass1lambda}-\eqref{ass1r}, \eqref{S1iterative}-\eqref{S3iterative}  with $\Sigma_{n-1}^{(1)} >0,\Sigma_{n-1}^{(2)} >0$ and $(\Sigma_{n-1}^{(3)}\big)^2 \le\Sigma_{n-1}^{(1)}\Sigma_{n-1}^{(2)}$, \eqref{eq:betaI}, the first part of \eqref{eq:betaRalpha}, as well as the second-order conditions \eqref{Isecond}-\eqref{Lsecond} hold. These are seven polynomial equations in seven unknown constants. We can then subsequently define $(\mu_n,s_n)$ by \eqref{ass1mu}-\eqref{ass1s} and $\alpha^R_n$ by the second part of \eqref{eq:betaRalpha}.

Next, the value function coefficients $( I_{n-1}^{(i,j)})_{1\leq i \leq j \leq 2}$ and $(L_{n-1}^{(i,j)})_{1\leq i \leq j \leq 3}$ at time $n-1$
are found by the recursions \eqref{I^{(1,1)}_n}-\eqref{I^{(2,2)}_n} and \eqref{L11}-\eqref{L33}.

\bigskip \noindent{\bf Termination:} The iteration above is continued back to time $n=0$.  If the resulting values at time $n=0$ satisfy \eqref{eq:starting-values}
 the algorithm terminates and the computed coefficients produce a linear Bayesian Nash equilibrium. Otherwise, we adjust the conjectured starting input values in \eqref{Sigma_N} and start the algorithm all over.

\section{Numerical results}
As is common with discrete-time Kyle-type models, we do not have analytic comparative results about the properties of our model.  However, we have conducted a variety of numerical experiments to illustrate properties of the model.  The baseline specification for our model has $N$ = 10 rounds of trading, the  variance of the terminal stock value $\tv$ is normalized to $\sigma^2_{\tilde v}$ = 1, the total variance of the Brownian motion noise trading order flow over the $N$ periods is fixed at $\sigma^2_{w}$ =  4, the variance of the trading target  $\ta$ is $\sigma^2_{\ta}$ = 1, and the correlation between the trading target $\ta$ and the terminal stock value $\tv$ is $\rho$ = 0 (i.e., $\tv$ and $\ta$ are ex ante independent). In our analysis, we  vary the correlation $\rho$ and the variance of the trading target $\sigma^2_{\ta}$.

\begin{figure}[!h]
\begin{center}
\caption{Plot of $(\lambda_n)_{n=1}^{N}$ for the parameters $\sigma^2_{\tv}=1$, $\sigma_w^2= 4$, $N=10$, $\sigma^2_{\ta}=1$ (right only), and $\rho =0$ (left only).}
$\begin{array}{cc}
\includegraphics[width=7cm, height=5cm]{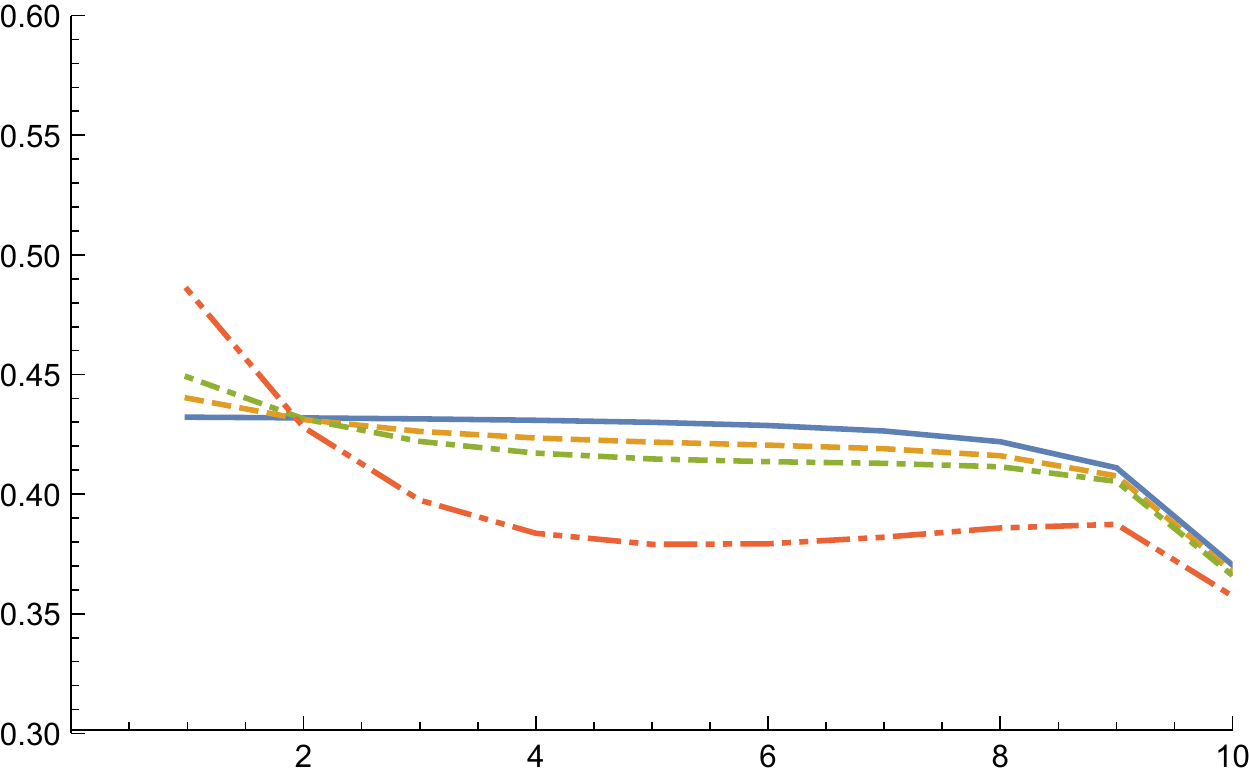}&
\includegraphics[width=7cm, height=5cm]{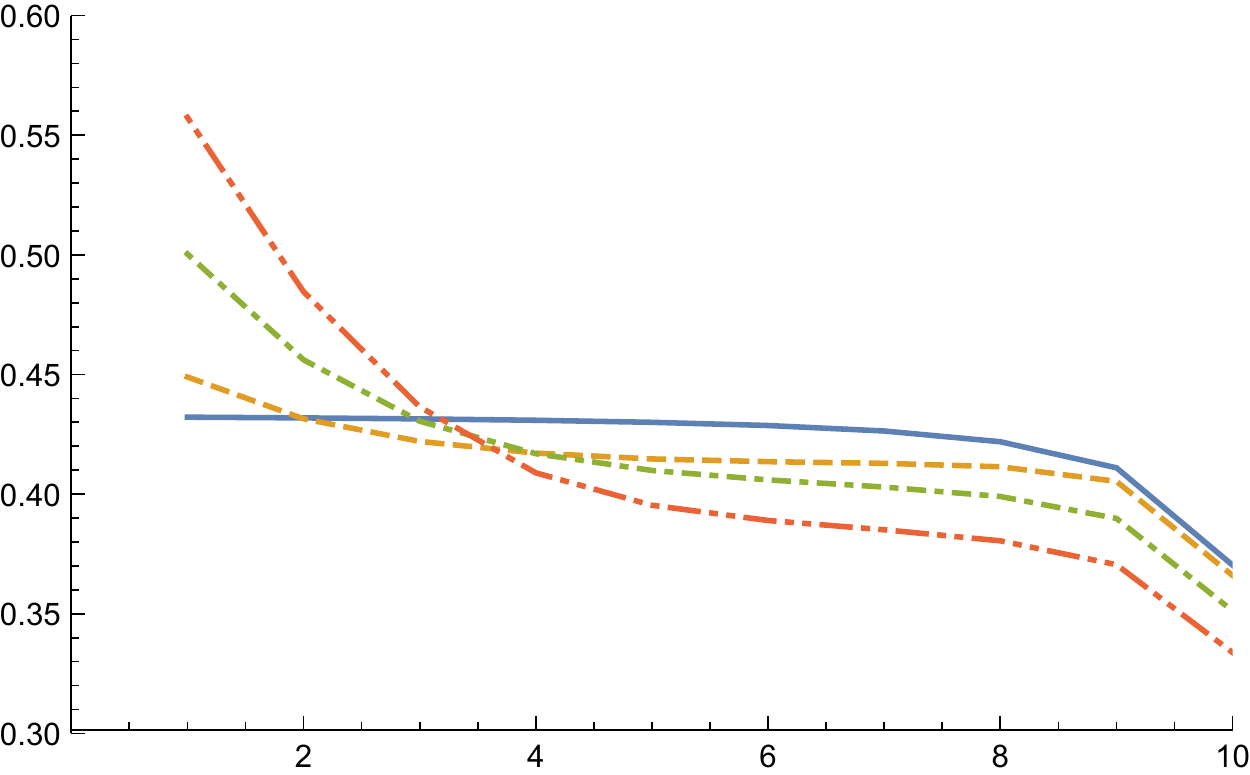}\\
 \text{A: Kyle (---------)}, \;\sigma^2_{\ta}=0.48\; (- - - ),& \text{B: Kyle  (---------)}, \;\rho=0\; (- - - ),\\
\sigma^2_{\ta}=1\;(-\cdot-\cdot-), \sigma^2_{\ta}=3.7\;(-\cdot\cdot-\cdot\cdot-).& \rho=0.25\;(-\cdot-\cdot-),  \rho=0.47\;(-\cdot\cdot-\cdot\cdot-).
\end{array}$
\label{fig:lambda}
\end{center}
\end{figure}

The two graphs in Figure \ref{fig:lambda} show the price impact of order flow parameter $\lambda_n$ over time.  The various dashed lines are for different parameterizations of our model. For comparison, the solid (blue) line is the corresponding price impact in Kyle (1985) in which the rebalancer is absent.  In the first round of trading at time $n = 1$, rebalancing noise by itself would reduce the value of $\lambda_1$ relative to Kyle (1985). However, in equilibrium, the insider's trading strategy also changes.  The net effect in this example is that $\lambda_1$ increases relative to Kyle (1985).\footnote{ We see in equation (\ref{ass1lambda}) that $\lambda_n$ is non-monotone in the aggressiveness of informed trading.  Thus, there may also be parameterizations for which our model has an inverted $U$-shape for $\lambda_n$.} At later times $ n >2$, the price impacts are lower than in Kyle. The result is an $S$-shaped twist in $\lambda_n$ over time. The price impact trajectory in our model also differs from Degryse, de Jong, and van Kervel (2014) in which price impacts have an inverted $U$-shape (see their Figure 1).

Figure \ref{fig:lambda}A varies the variance of the trading target $\sigma^2_{\ta}$ while holding $\rho$ fixed at 0.  We see that the $S$-shaped twist in $\lambda_n$ becomes stronger for larger values of $\sigma^2_{\ta}$. When $\sigma^2_{\ta}$ is high enough, the price impact of order flow can even be non-monotone over time (see the dashed line corresponding to $\sigma^2_{\ta} = 3.7$, which is comparable to the total daily noise trader order variance  $\sigma_w^2= 4$). Figure \ref{fig:lambda}B varies the correlation $\rho$ between the terminal stock value $\tv$ and the trading target $\ta$ while holding the variance $\sigma^2_{\ta}$ fixed at 1.  Here again, there is an asymmetric impact of $\rho$ over time relative to our baseline model with $\rho = 0$. At early times, $\lambda_n$ is increasing in the correlation $\rho$, but at later times, $\lambda_n$ is decreasing in $\rho$. This is because increasing $\rho$ changes some rebalancing trades from noise into informative order flow.

\begin{figure}[!h]
\begin{center}
\caption{Plot of $(\Sigma^{(2)}_n)_{n=1}^{N}$ for the parameters $\sigma^2_{\tv}=1$, $\sigma_w^2= 4$, $N=10$, $\sigma^2_{\ta}=1$ (right only), and $\rho =0$ (left only).}
$\begin{array}{cc}
\includegraphics[width=7cm, height=5cm]{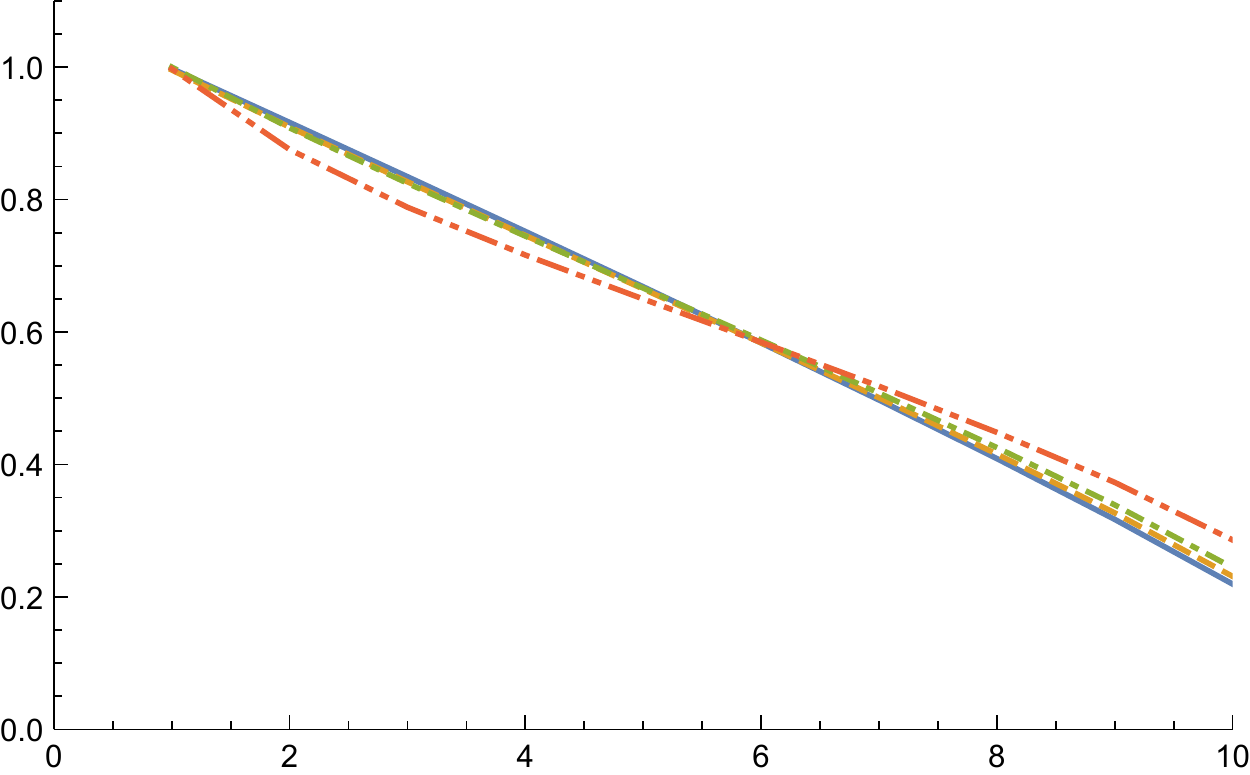} &
\includegraphics[width=7cm, height=5cm]{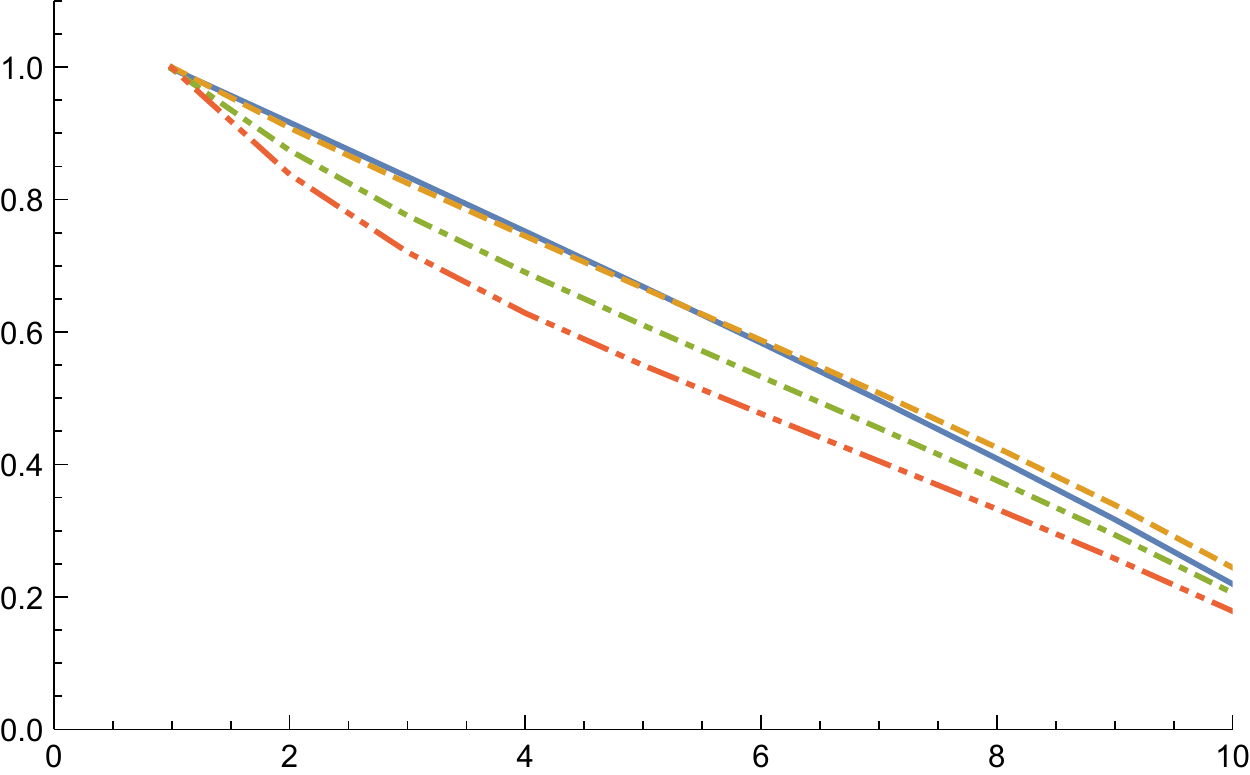}
\\
 \text{A: Kyle (---------)}, \;\sigma^2_{\ta}=0.48\; (- - - ),& \text{B: Kyle  (---------)}, \;\rho=0\; (- - - ),\\
\sigma^2_{\ta}=1\;(-\cdot-\cdot-), \sigma^2_{\ta}=3.7\;(-\cdot\cdot-\cdot\cdot-).& \rho=0.25\;(-\cdot-\cdot-),  \rho=0.47\;(-\cdot\cdot-\cdot\cdot-).
\end{array}$
\label{fig:sigma2}
\end{center}
\end{figure}

Figure \ref{fig:sigma2} shows the trajectory of the variance $\Sigma_n^{(2)}$ of $\tv -p_{n-1}$ over time where $(p_n)_{n=1}^N$ are the equilibrium prices.  In our baseline case where $\rho = 0$, there is faster information revelation at early times, but slower information revelation later towards the end. When $\rho > 0$, public uncertainty about $\tv$ falls faster in our model than in Kyle's model. This is because, with $\rho > 0$, the rebalancer also trades, from the beginning, on information about the stock value.

\begin{figure}[!h]
\begin{center}
\caption{Plot of $(\beta^I_n)_{n=1}^{N}$ for the parameters $\sigma^2_{\tv}=1$, $\sigma_w^2= 4$, $N=10$, $\sigma^2_{\ta}=1$ (right only), and $\rho =0$ (left only).}
$\begin{array}{cc}
\includegraphics[width=7cm, height=5cm]{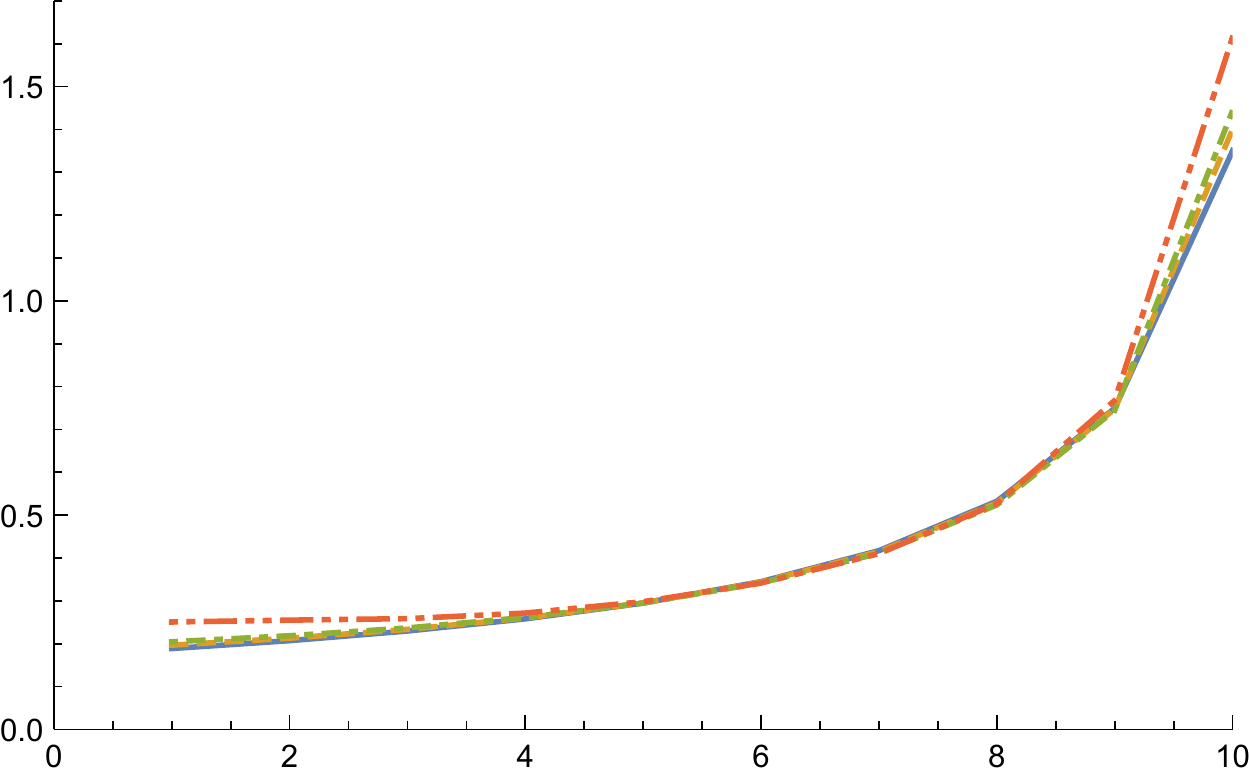}&
\includegraphics[width=7cm, height=5cm]{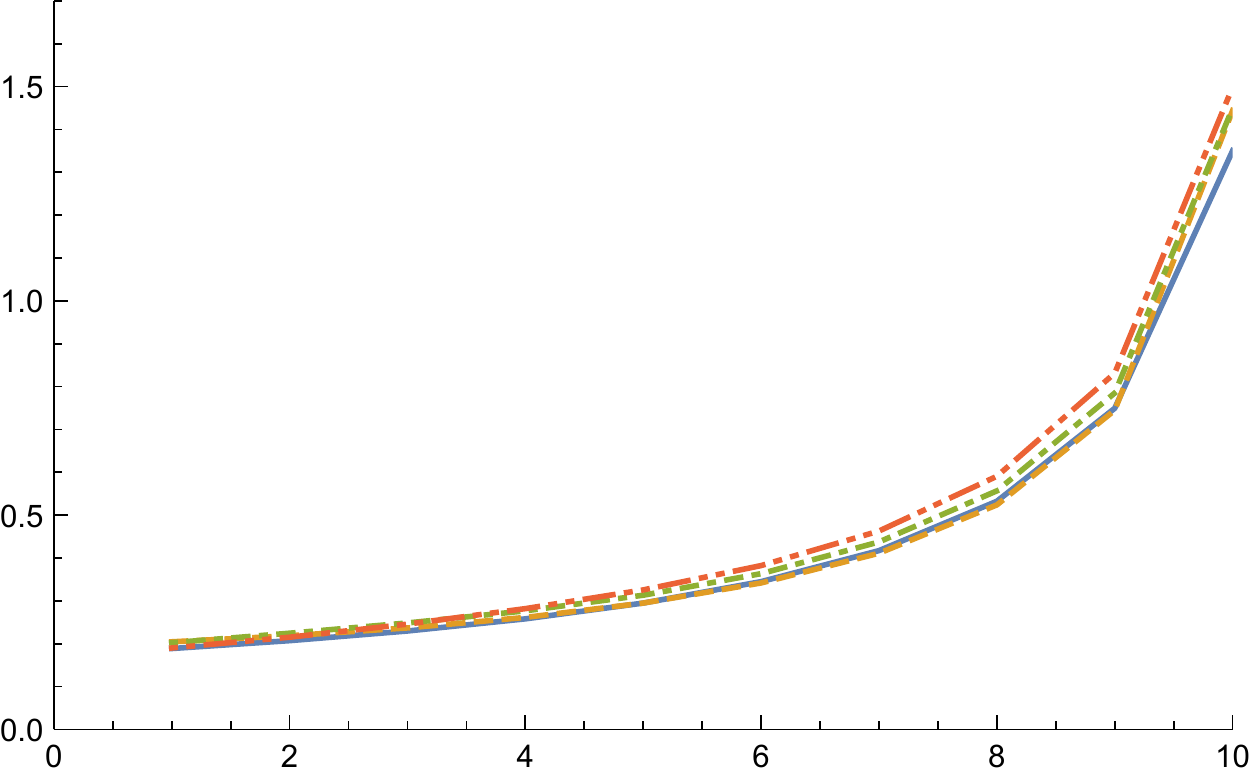}
\\
\text{A: Kyle (---------)}, \;\sigma^2_{\ta}=0.48\; (- - - ),& \text{B: Kyle  (---------)}, \;\rho=0\; (- - - ),\\
\sigma^2_{\ta}=1\;(-\cdot-\cdot-), \sigma^2_{\ta}=3.7\;(-\cdot\cdot-\cdot\cdot-).& \rho=0.25\;(-\cdot-\cdot-),  \rho=0.47\;(-\cdot\cdot-\cdot\cdot-).
\end{array}$
\label{fig:betaI}
\end{center}
\end{figure}

Figure \ref{fig:betaI} shows the insider's strategy coefficients $\beta^I_n$, which measures how aggressively she trades on her private information $\tv - p_{n-1}$ over time.  As in Kyle, the intensity of informed trading in our model increases as time approaches the terminal time $N$. This is consistent with the fact that the price impact of order flow $\lambda_n$ in Figure \ref{fig:lambda} shrinks as time passes.  We also see that as the variance  of the trading target $\sigma^2_{\ta}$ increases, the informed investor trades more aggressively at early dates, less so in the middle, and then slightly more aggressively again towards the end. The informed trader's increased initial aggressiveness reflects the fact that there is more noise, due to the rebalancer's trading target $\ta$, in which to hide the insider's orders. In addition, if $\rho > 0$, insider trading aggressiveness increases somewhat due to a Holden-Subrahmanyam race--to--trade competition effect. The apparent size of the changes in $\beta^I_1$ -- which are on the order of 10 percent  -- are visually understated in Figure \ref{fig:betaI} because of the vertical scaling (due to the size of $\beta_{10}^I$).\footnote{  With $\rho > 0$, there are two differences relative to Holden and Subrahmanyam (1992).  First, the insider still has better information than the rebalancer if $\rho < 1$.  Thus, our analysis with $\rho > 0$ is more comparable to Foster and Viswanathan (1994), which has two asymmetrically informed traders, one of which is better informed than the other. Second, trading by our rebalancer, when he is informed about $\tv$, is constrained by his terminal target $\ta$.  This works against rat races with extremely aggressive rebalancer trading. }

\begin{figure}[!h]
\begin{center}
\caption{Plot of $\E[\Delta\theta^I_n|\sigma(\tv)]$ for $n=1,2,...,10$.  The parameters are $\sigma^2_{\tv}=1$, $\sigma_w^2= 4$, $N=10$, $\sigma^2_{\ta}=1$ (right only), $\rho =0$ (left only), and the realization of $\tv$ equals 1.}
$\begin{array}{cc}
\includegraphics[width=7cm, height=5cm]{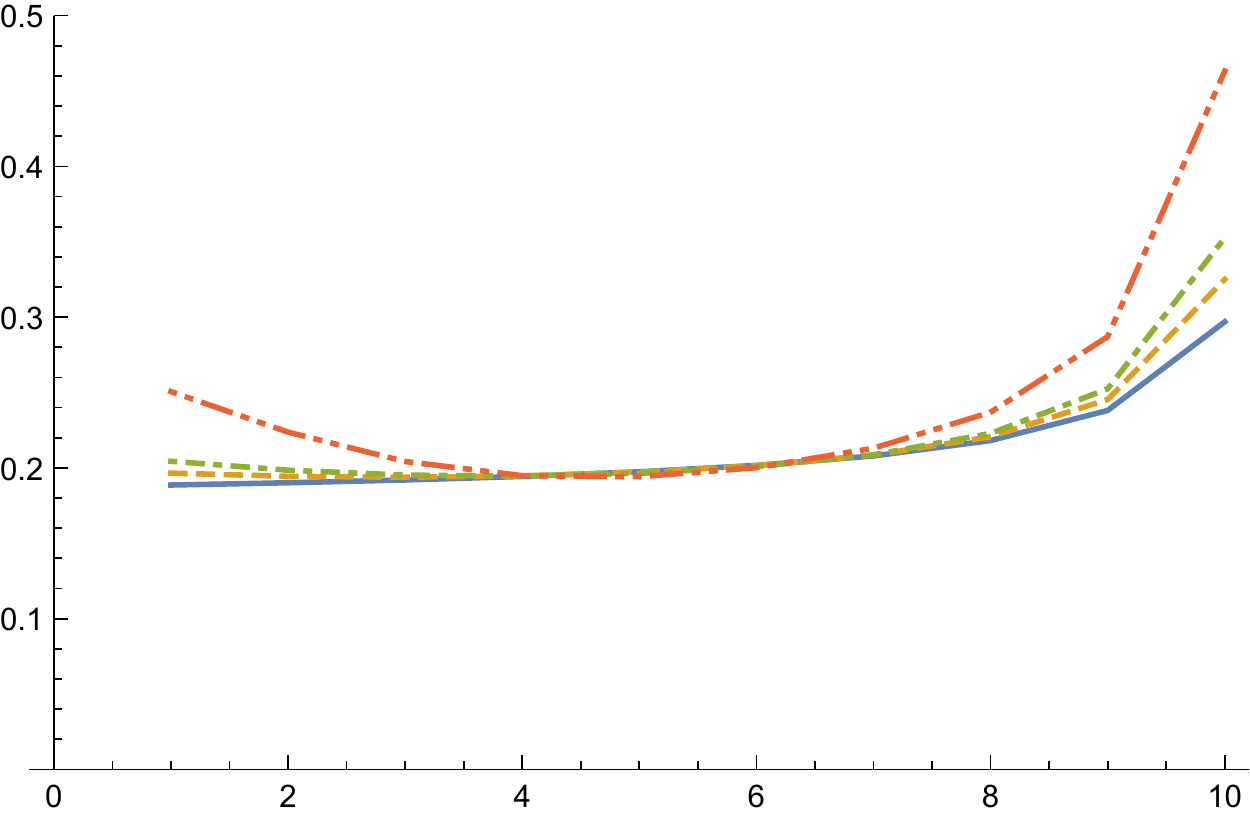} &\includegraphics[width=7cm, height=5cm]{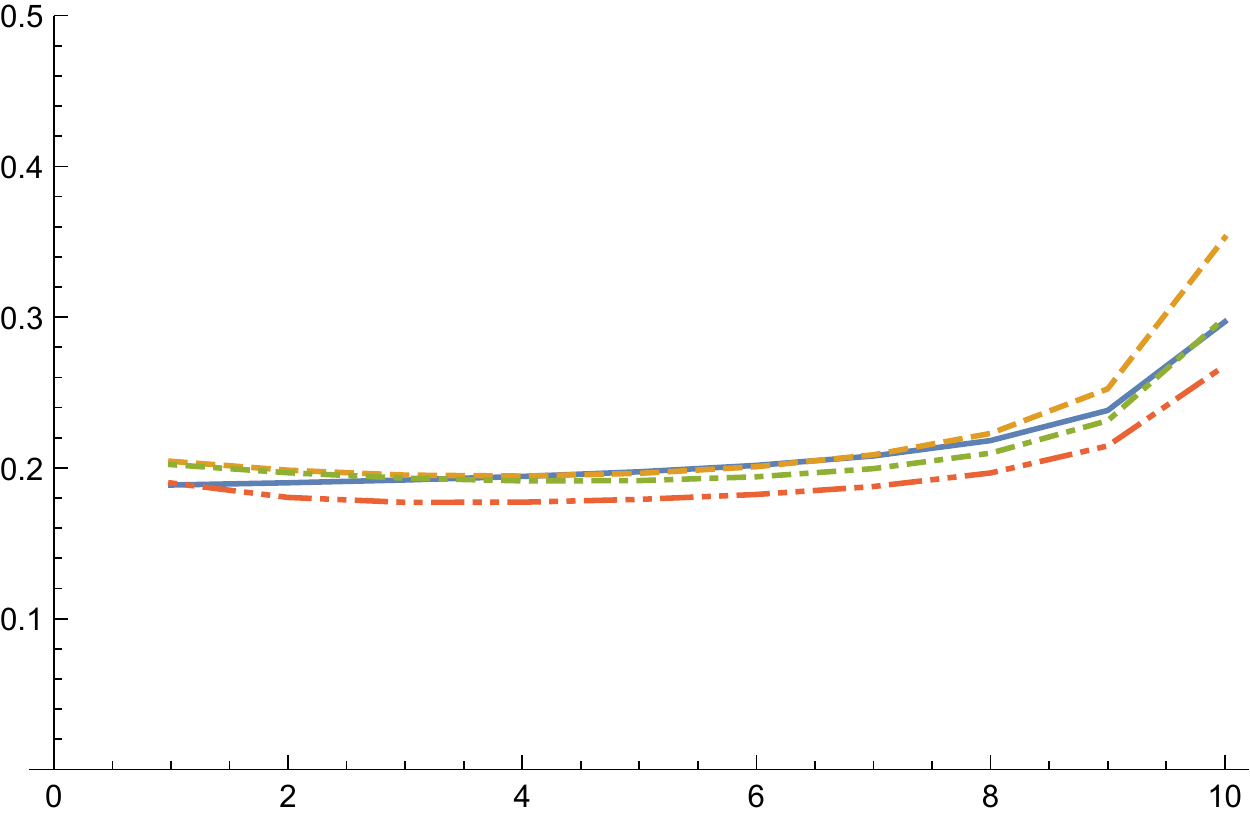}
\\
\text{A: Kyle (---------)}, \;\sigma^2_{\ta}=0.48\; (- - - ),& \text{B: Kyle  (---------)}, \;\rho=0\; (- - - ),\\
\sigma^2_{\ta}=1\;(-\cdot-\cdot-), \sigma^2_{\ta}=3.7\;(-\cdot\cdot-\cdot\cdot-).& \rho=0.25\;(-\cdot-\cdot-),  \rho=0.47\;(-\cdot\cdot-\cdot\cdot-).
\end{array}$
\label{fig:dthetaI}
\end{center}
\end{figure}

Figure \ref{fig:dthetaI} shows the insider's expected trades (i.e., conditional on her information) over the day for the specific value realization $\tv = 1$ and averaged over $\ta$ and noise trader paths $w$. Kyle's model is the solid (blue) line, whereas the dotted lines represent various parameterizations of our model. Unlike Kyle's model, our model produces a slight $U$-shaped trading pattern; that is, our insider expects ex ante to trade somewhat more initially and again at the end of the day. However, the $U$-shape is not big. Since the trading expectations in Figure \ref{fig:dthetaI} are linear in the realization of $\tv$, the expected informed trading volume is also slightly $U$-shaped for other realizations of $\tv$.

Next, we turn to the rebalancer. The rebalancer's trades reflect a variety of considerations:  First, the rebalancer needs to reach his trading target $\ta$ at time $N$.  Second, he wants to reach this target at the lowest cost possible.  Thus, to the extent that his orders have a price impact, he splits up his orders to take into account the pattern of the price impact coefficients $\lambda_n$ over time. Third, the rebalancer engages in ``sunshine trading." In particular, early orders signal predictable future orders at later dates, which, from \eqref{eq_pp}, will have no price impact. Fourth, the rebalancer understands that price pressure from his trades creates incentives for the insider to trade.  At later dates, this can actually be beneficial for the rebalancer. For example, if early uninformed rebalancer orders raise prices, then, in expectation, the insider should then buy less/sell more in the future, thereby putting downward pressure on later prices which, in turn, reduces the expected cost of subsequent rebalancer buying. Fifth, the rebalancer trades on information about the asset value $\tv$.  If $\rho > 0$, the rebalancer starts out with stock valuation information.  However, even if the rebalancer is initially uninformed about $\tv$ (i.e., $\rho=0$), he still acquires stock valuation information over time (see \ref{eq:rebalancer-information-from-trading}) that he can use to reduce his rebalancing costs and even, possibly, to earn a trading profit. In particular, he can filter the aggregate order flow to learn about the insider's trading, and thereby learn about $\tv$, better than the market makers.


To gain further intuition, we rearrange (\ref{theta_L}) to decompose the rebalancer's order at time $n$ as follows:
\begin{equation}\label{eq:RabOrderDecomposition}
\Delta\theta^R_n = \beta_n^R (\ta - \theta^R_{n-1 } - q_{n-1}) + (\alpha_n^R + \beta_n^R) q_{n-1}.
\end{equation}
From (\ref{hat_qn}), second component, $(\alpha_n^R + \beta_n^R) q_{n-1}$, is the market makers' expectation $\E[\Delta \theta_n^R | \sigma(y_1, \ldots, y_{n-1})]$ of the rebalancer's
order at time $n$.  This amount is traded at time $n$ with no price impact.  The
first component, $\beta_n^R (\ta - \theta^R_{n-1} - q_{n-1})$, represents the
combined effect of i) strategic trading by the rebalancer on his private
information, $\ta - \theta^R_{n-1} - q_{n-1}$, which is informative about
$\tv-p_{n-1}$ (see \ref{eq:rebalancer-information-from-trading}), and ii) rebalancing trading
given that the remaining amount that the rebalancer actually needs to
trade (i.e.,  $\ta - \theta^R_{n-1 })$ differs, in general, from the market makers'
expectation $q_{n-1}$.

\begin{figure}[!h]
\begin{center}
\caption{Plot of $(\alpha^R_n)_{n=1}^N$ (below the $x$-axis) and of $(\beta^R_n)_{n=1}^N$ (above the $x$-axis) for $n=1,2,...,10$.  The parameters are $\sigma^2_{\tv}=1$, $\sigma_w^2= 4$, $N=10$, $\sigma^2_{\ta}=1$ (right only), and $\rho =0$ (left only).}
$\begin{array}{cc}
\includegraphics[width=7cm, height=5cm]{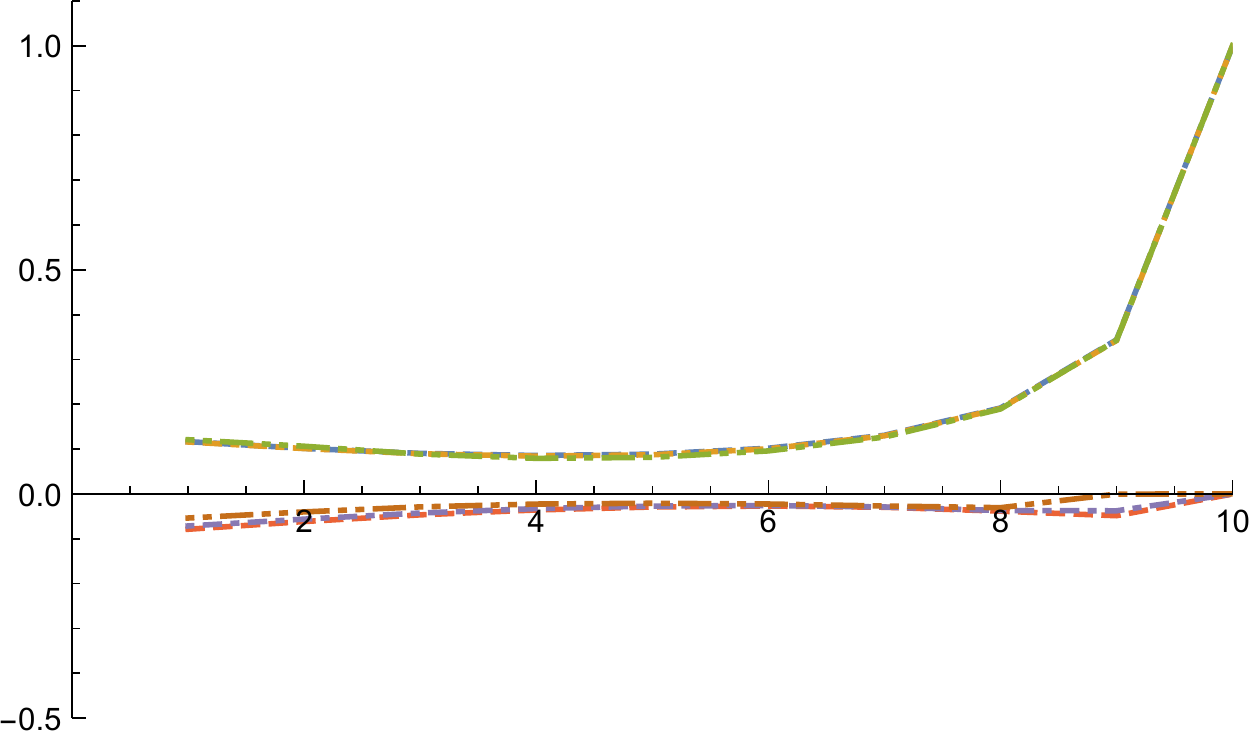} &
\includegraphics[width=7cm, height=5cm]{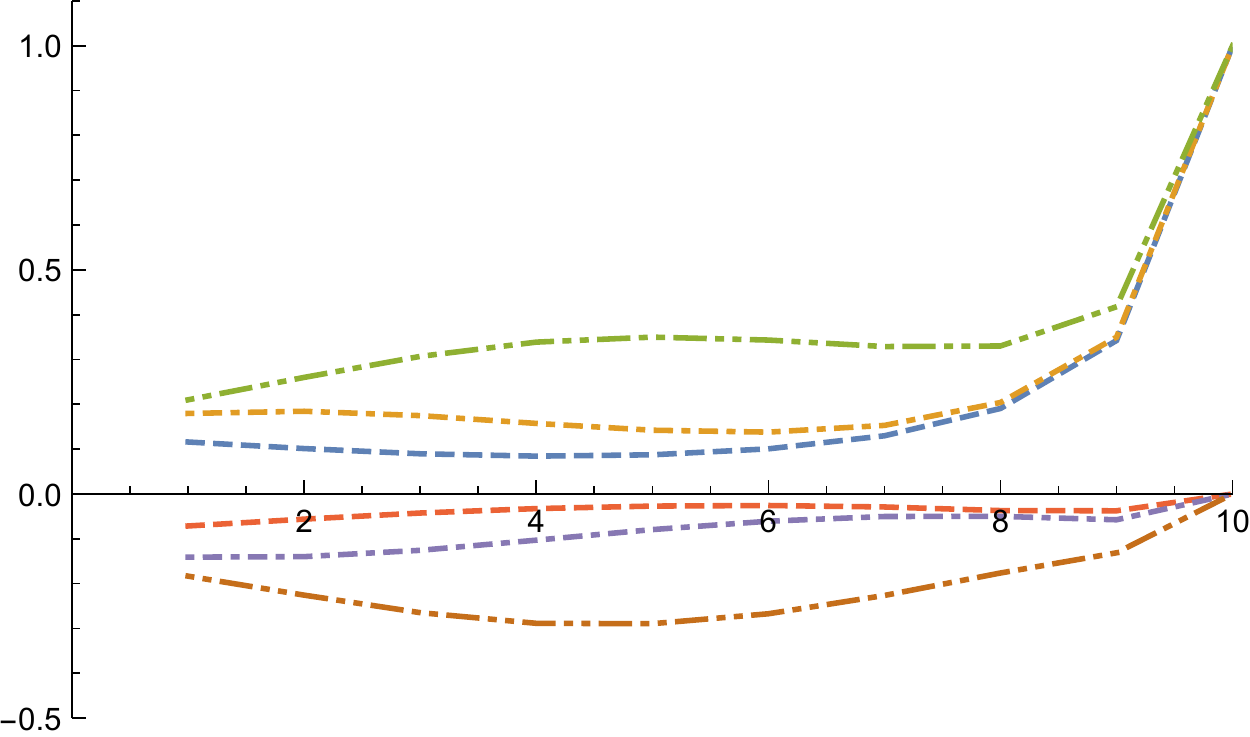}
\\
   \text{A: }\sigma^2_{\ta}=0.48\; (- - - ),\;\sigma^2_{\ta}=1\;(-\cdot-\cdot-), &\text{B: } \rho=0\; (- - - ),\;\rho=0.25\;(-\cdot-\cdot-),\\
  \sigma^2_{\ta}=3.7\;(-\cdot\cdot-\cdot\cdot-).   &\rho=0.47\;(-\cdot\cdot-\cdot\cdot-).
\end{array}$
\label{fig:alphaLbetaL}
\end{center}
\end{figure}

Figure \ref{fig:alphaLbetaL} shows trajectories for the rebalancer's strategy coefficients $\beta_n^R$ and $\alpha_n^R$. We use the decomposition \eqref{eq:RabOrderDecomposition} to interpret them.  Since $\alpha_n^R + \beta_n^R $ is positive but
small until time $N$, the rebalancer trades a relatively small
fraction of his expected trading gap $q_{n-1}$ over time until time $N$ at
which time $\alpha_N^R + \beta_N^R  = 1$ and then he trades the full remaining
gap.  In addition, the fact that $\beta_n^R$ is
positive means that the rebalancer trades in the direction of his private
information.  He does this for two reasons: First, the larger $\ta$ is
relative to  $\theta^R_{n-1 } $ (given $q_{n-1}$), the more the rebalancer
must trade to achieve his target compared to the market makers'
expectation of his trading gap.  Second, the smaller $\theta_{n-1}^R$ is relative to
$q_{n-1}$ (given $\ta$), the less the rebalancer has actually bought relative
to the market makers' expectation, which, in turn, implies that, given the
prior observed aggregate order flows, the more the insider bought in
expectation given the rebalancer's information. Hence, in this situation the rebalancer infers that the market makers have, on average, underpriced the stock
and, therefore, strategically buys more/sells less stock at times $n < N$.

\begin{figure}[!h]
\begin{center}
\caption{Plot of $\E[\Delta\theta^R_n|\sigma(\ta)]$ for $n=1,2,...,10$.  The parameters are $\sigma^2_{\tv}=1$, $\sigma_w^2= 4$, $N=10$, $\sigma^2_{\ta}=1$ (right only), $\rho =0$ (left only), and the realization of $\ta$ equals 1.}
$\begin{array}{cc}
\includegraphics[width=7cm, height=5cm]{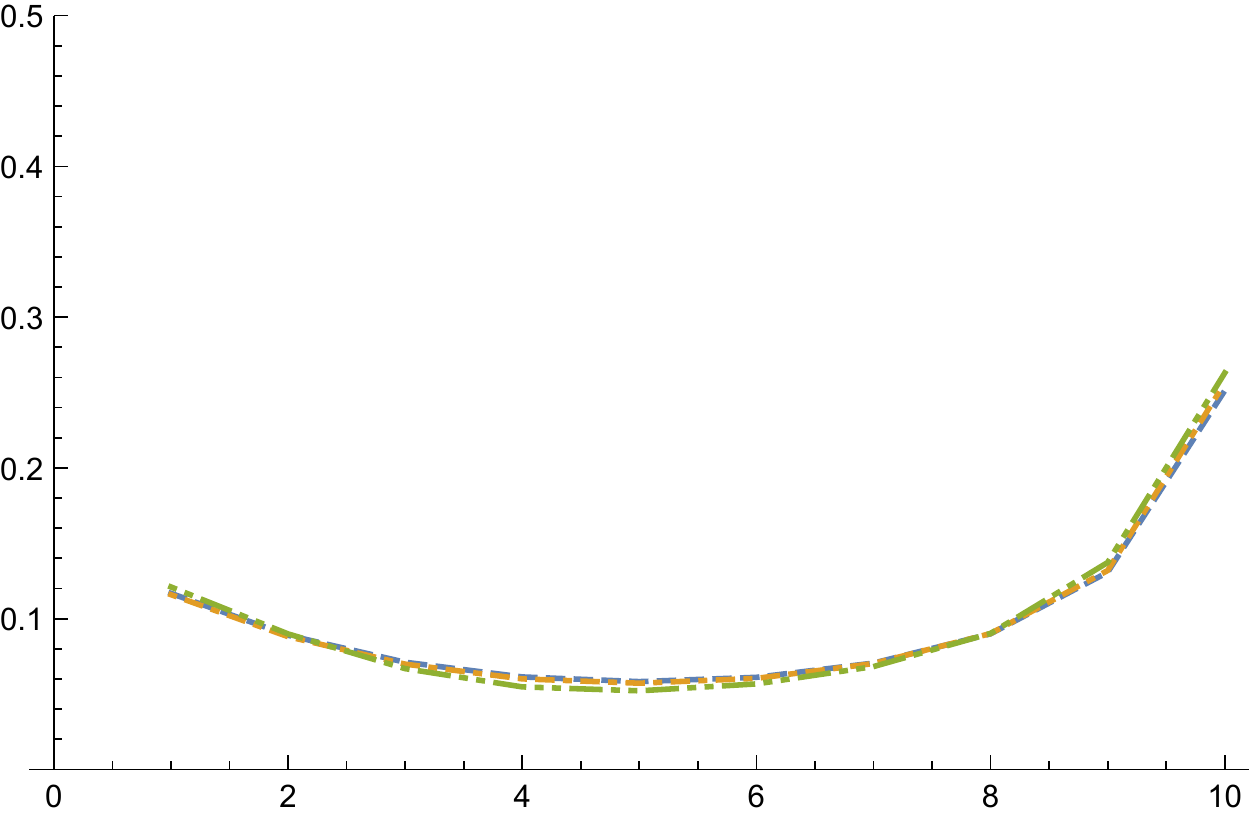}&\includegraphics[width=7cm, height=5cm]{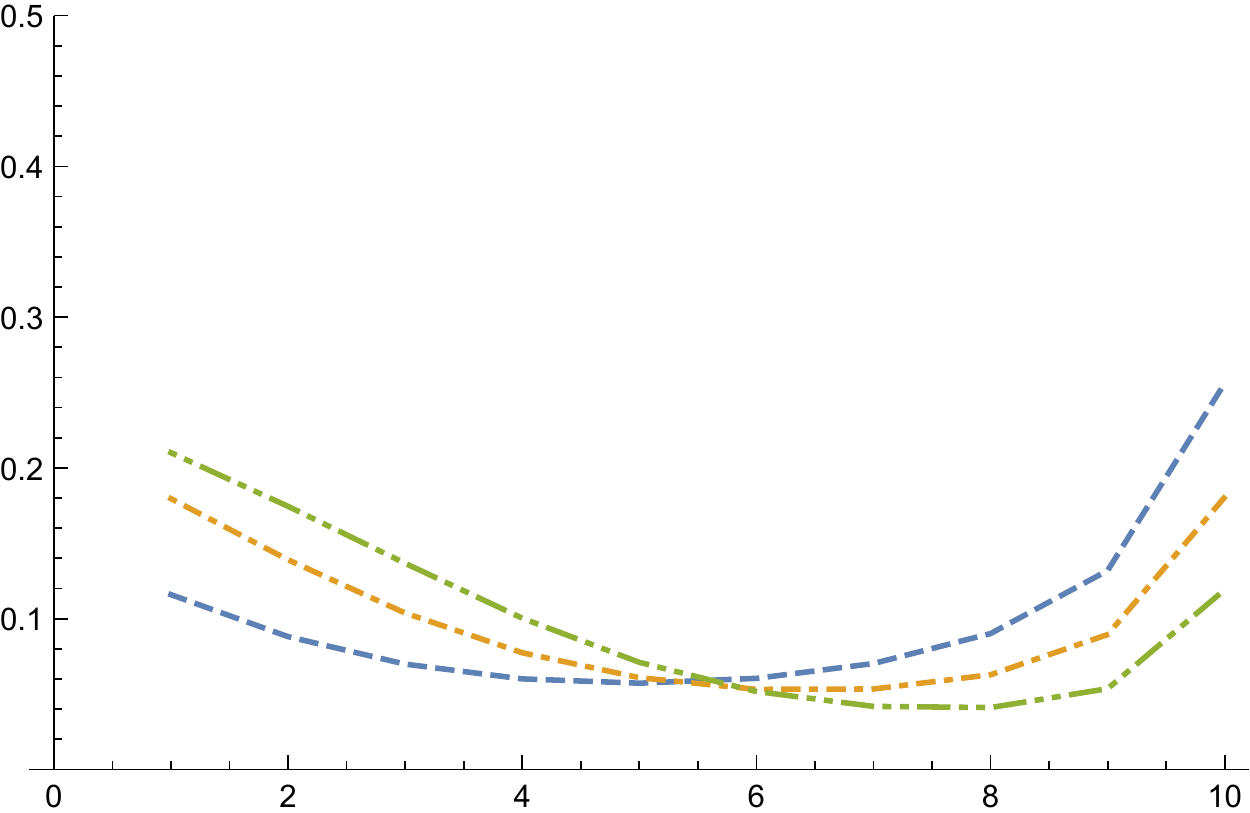}
\\
   \text{A: }\sigma^2_{\ta}=0.48\; (- - - ),\;\sigma^2_{\ta}=1\;(-\cdot-\cdot-), &\text{B: } \rho=0\; (- - - ),\;\rho=0.25\;(-\cdot-\cdot-),\\
  \sigma^2_{\ta}=3.7\;(-\cdot\cdot-\cdot\cdot-).   &\rho=0.47\;(-\cdot\cdot-\cdot\cdot-).
\end{array}$
\label{fig:expectdthetaL}
\end{center}
\end{figure}

Figure \ref{fig:expectdthetaL} shows the rebalancer's ex ante expected orders over the day for the particular realization of the trading target $\ta$ being equal to 1. These expectations are taken over the terminal stock price $\tv$ and the noise trader order path $w$. These expectations depend linearly on the realization of the trading target $\ta$. The graphs show that the rebalancer's trading strategy also has a $U$--shaped pattern over the day.  Degryse, de Jong, and van Kervel (2014) obtain a similar result in their model with short--lived information for the insiders and static trading for the rebalancer. In particular, with short-lived information, their insider is unable to trade dynamically over time, which allows the rebalancer to (imperfectly) separate his order from those of the insider.  In contrast, in our model, the insider trades dynamically too.  Thus, the $U$-shaped pattern of rebalancing trading does not depend on the assumption of short-lived information.

\begin{figure}[!h]
\begin{center}
\caption{Properties of the rebalancer's orders. The parameters are $N:=10$, $\sigma_w^2 := 4$, $\sigma_{\tv}^2:=1$,  $\sigma^2_{\ta} :=1$, and $\rho :=0$.}
\vspace{.25cm}
$\begin{array}{cc}
\includegraphics[width=7cm, height=5cm]{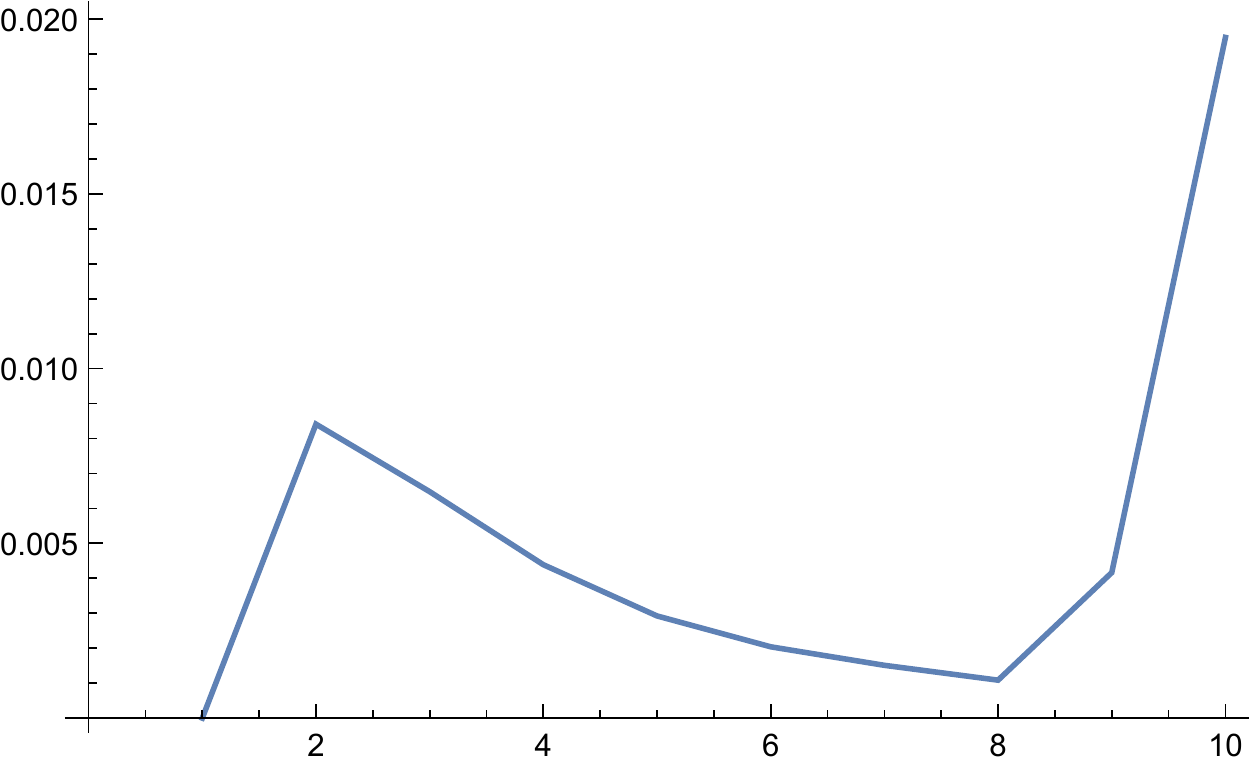}&
\includegraphics[width=7cm, height=5cm]{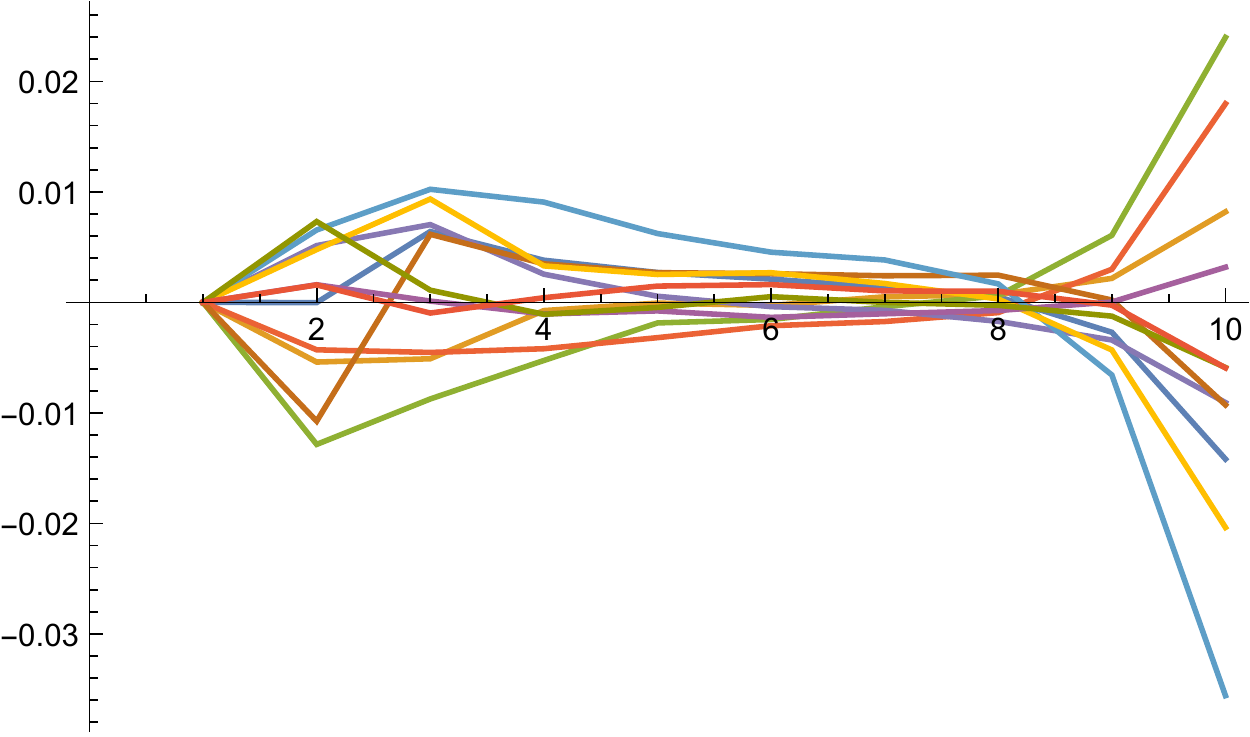}
\\
\text{A: }\E\big[ \big(\Delta \theta_n^R - \E[\Delta \theta_n^R|\sigma(\ta)]\big)^2\big|\sigma(\ta)\big]^\frac12, &\text{B: } \text{10 paths of}\; \Delta \theta_n^R \\
\text{for the realization } \ta =1. & \text{ for the realizations $\ta =0$ and $\tv=1$.}
\end{array}$

\label{fig:10paths}
\end{center}
\end{figure}

The literature on optimal order execution includes many models that also produce $U$-shaped optimal strategies, see, e.g., Predoiu, Shaikhet, and Shreve (2011) and the references therein. However, sunshine trading in that literature stems from exogenously specified liquidity resilience and replenishment dynamics. In contrast, liquidity in our equilibrium model is endogenously determined. In our model, there are two sources of $U$-shaped rebalancer trading volume.  First, orders from the rebalancer early in the day signal to the market makers the size of the predictable component of his orders at the end of the day. Second, there are also $U$-shaped patterns in the standard deviation of rebalancer orders. In particular, because the rebalancer's trades depend on the aggregated order flow history via $q_n$, there is variability across the rebalancer's order flow paths. Figure \ref{fig:10paths}A shows the ex ante standard deviation of the rebalancer's orders over the day conditional on the rebalancer's target $\ta$. Here again, we see a $U$-shaped pattern.

Figure \ref{fig:10paths}B plots a few paths of the rebalancer's order flows over time. Here the realized stock value $\tv$ is 1, and the realized trading target $\ta$ is 0. There are 10 different randomly selected path realizations of the noise traders' orders.  Along these paths, we see that the rebalancer buys/sells  more than his trading target $\ta$ at early dates ($n > 1$) and then unwinds his position at later dates to achieve his trading target.  This is not manipulation.  Rather, the rebalancer's orders reflect a combination of informed trading motives (about $\tv$) and uninformed rebalancing motives (due to $\ta$).  The rebalancer does not trade at time 1 because he does not need to rebalance and because, initially, he does not have any stock valuation information (i.e., $\rho=0$).  However, at time 2 the rebalancer trades based on whether --- given the value information he gleans from being able to filter the order flow $y_1$ better than the market makers -- he thinks the stock is over-- or under--valued.  Eventually, however, he must unwind these earlier positions in order to achieve his realized trading target constraint $\theta^R_N = \ta = 0$ at the end of the day.\footnote{This is another example of a situation in which different traders acquire information at different times and/or have to unwind positions in advance of definitive public announcements. See also Foucault, Hombert, and Rosu (2015).} The dispersion in the paths is consistent with the trajectory of the rebalancer order flow standard deviation.  Paths for non-zero values of $\ta$ involve shifting the means of these paths from zero to the appropriate ex ante conditional means given $\ta$ (e.g., Figure \ref{fig:expectdthetaL} illustrates one such conditional mean order flow trajectory for $\ta = 1$).

\begin{figure}[!h]
\begin{center}
\caption{Plot of $\frac{ \E[y_ny_{n+1}]}{\sqrt{\E[y_n^2]\E[y_{n+1}^2]}}$ for $n=1,2,...,9$.  The parameters are $N:=10$, $\sigma_w^2 := 4$, $\sigma_{\tv}^2:=1$,  $\sigma^2_{\ta} =1$ (right only), and $\rho =0$ (left only).}
$\begin{array}{cc}
\includegraphics[width=7cm, height=5cm]{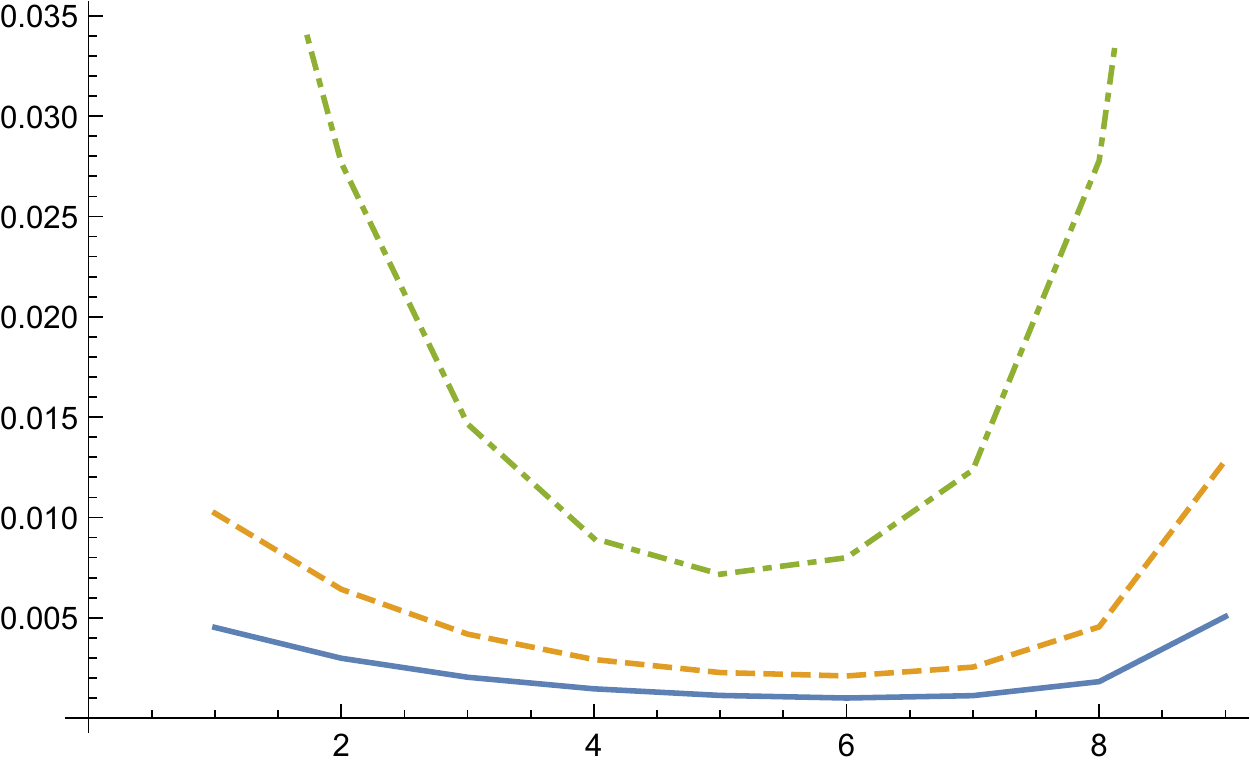} &
\includegraphics[width=7cm, height=5cm]{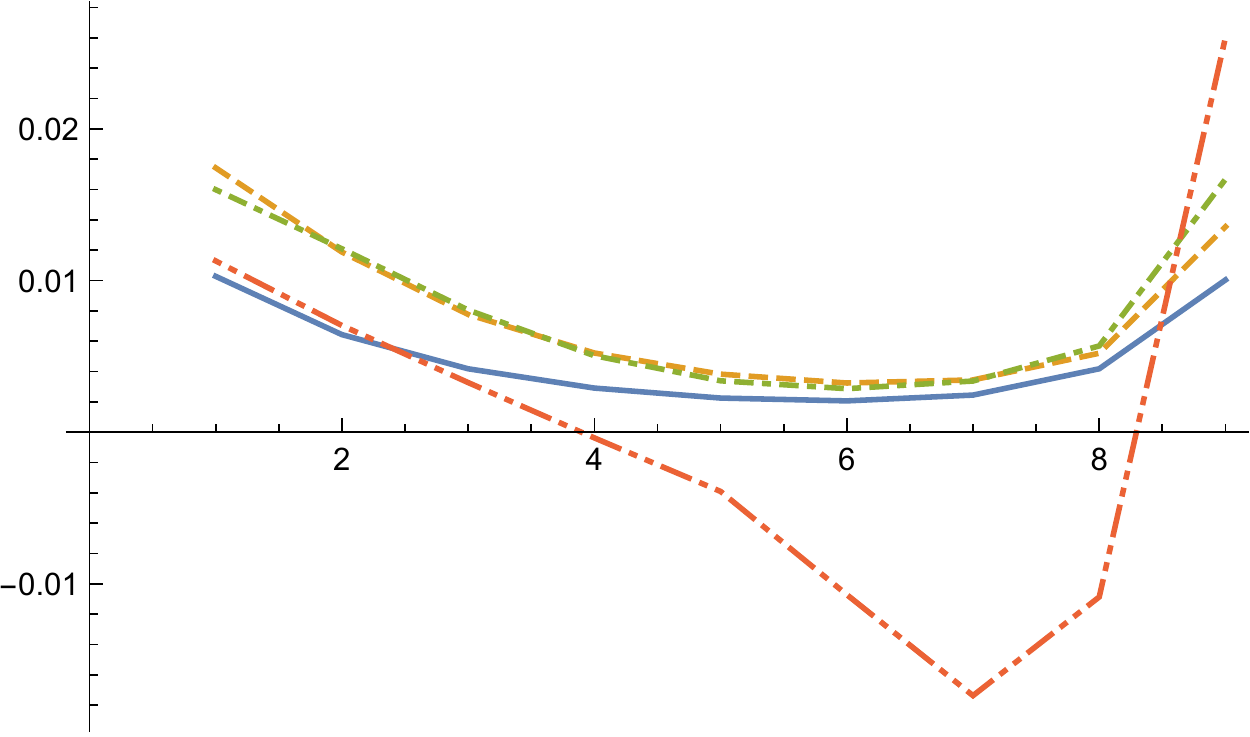}
\\
\text{A: }\sigma_{\ta}^2 =0.48\; \text{(---------)}, \,\sigma_{\ta}^2 = \;1\;(- - - ),&\text{B: }\rho =0 \text{(---------)}, \,\rho = .25\;(- - - ),\\  \sigma_{\ta}^2 = 3.7\; (-\cdot-\cdot-).&\rho = .47 \;(-\cdot-\cdot-),\; \rho = .86\; (-\cdot\cdot-\cdot\cdot-).
\end{array}$
\label{dec8_1}
\end{center}
\end{figure}

Figure \ref{dec8_1} shows the unconditional autocorrelation of the aggregate order flow over time for different values of $\sigma^2_{\ta}$ and $\rho$.  Although the absolute level of autocorrelation is low, there is a clear $U$-shaped pattern of higher order flow autocorrelation at the beginning and the end of the day (when, from Figure  \ref{fig:expectdthetaL}, the rebalancer is trading more) with lower autocorrelation during the middle of the day (when the rebalancer trades less).  Somewhat surprisingly, order flow autocorrelation can be negative in the middle of the day when the target-information correlation $\rho$ is high.

\begin{figure}[!h]
\begin{center}
\caption{Plot of $\sqrt{\E[(p_n-p_{n-1})^2]}$ for $n=1,2,...,10$.  The parameters are $\sigma^2_{\tv}=1$, $\sigma_w^2= 4$, $N=10$, $\sigma^2_{\ta}=1$ (right only), and  $\rho =0$ (left only).}
$\begin{array}{cc}
\includegraphics[width=7cm, height=5cm]{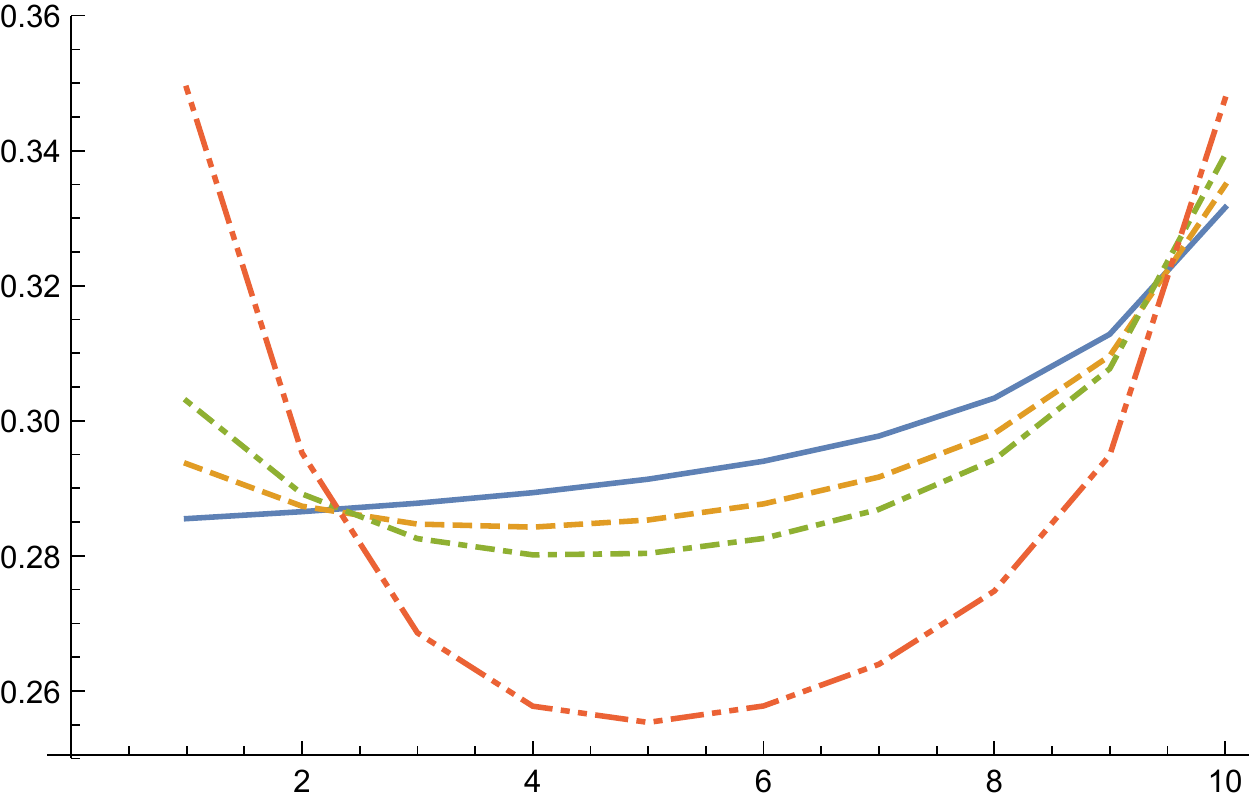} &\includegraphics[width=7cm, height=5cm]{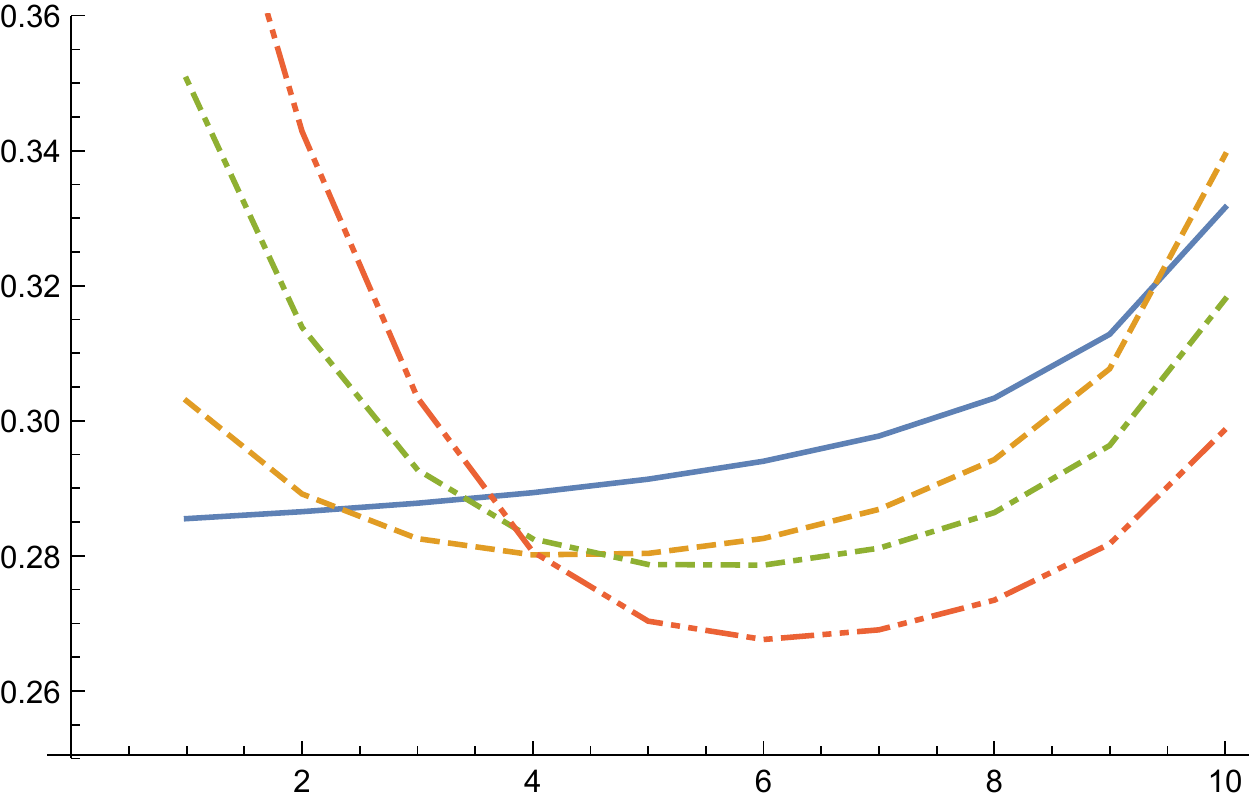}
\\
\text{A: Kyle (---------)}, \;\sigma^2_{\ta}=0.48\; (- - - ),&\text{B: Kyle  (---------)}, \;\rho=0\; (- - - ),\\
\sigma^2_{\ta}=1\;(-\cdot-\cdot-), \sigma^2_{\ta}=3.7\;(-\cdot\cdot-\cdot\cdot-).& \rho=0.25\;(-\cdot-\cdot-),  \rho=0.47\;(-\cdot\cdot-\cdot\cdot-).
\end{array}$
\label{fig:dp_sd}
\end{center}
\end{figure}

Figure \ref{fig:dp_sd} shows the unconditional standard deviation for the price changes over time. Kyle's model is the solid (blue) line, which is monotonically increasing, whereas our model produces the $U$-shaped dotted lines (for various correlation parameters $\rho$ and target variances $\sigma^2_{\ta}$). In other words, our model produces equilibrium prices which are more volatile at the beginning and at the end of the trading day relative to the middle of the trading day.

The rebalancer's trading strategy takes into account two types of predictability in his orders.  One part of his orders is predictable to the market makers based on the prior aggregate order flow. As mentioned after \eqref{eq:RabOrderDecomposition}, the rebalancer's sunshine trading component (i.e., the part that is predictable for the market makers) of his order at time $n$ is $(\beta^R_n+ \alpha^R_n) q_{n-1}$.
The advantage to the rebalancer of sunshine trading predictability is that this part of his trades has no price impact (see \ref{eq_pp}).

Another part of the rebalancer's orders is predictable to the insider. In particular, as shown in \eqref{eq:insider-information}, the insider can filter the aggregate order flow better than the market makers to identify rebalancing orders.  The part that is predictable to the insider is
\begin{align}
\begin{split}
\E[&\Delta \theta_n^R | \sigma(\tv,y_1, \ldots, y_{n-1})]\\ &=\beta_{n}^R\E[ (\tilde{a}-\theta_{n-1}^R) | \sigma(\tv,y_1, \ldots, y_{n-1})] +\alpha_n^R q_{n-1} \\&=
\beta_{n}^R\E[ (\tilde{a}-\theta_{n-1}^R-q_{n-1}) | \sigma(\tv-p_{n-1})] +(\alpha_n^R+\beta^R_n) q_{n-1} \\
&=\beta_{n}^R\frac{\Sigma^{(3)}_n}{\Sigma^{(2)}_n}(\tv-p_{n-1})+(\alpha_n^R+\beta^R_n) q_{n-1}.
\end{split}
\end{align}
Consequently, the part of the rebalancer's order $\Delta\theta^R_n$ which is expected by the insider (and not the market makers) is given by
\begin{align}
\E[\Delta \theta_n^R | \sigma(\tv,y_1, \ldots, y_{n-1})]-\E[\Delta \theta_n^R | \sigma(y_1, \ldots, y_{n-1})]=\beta_{n}^R\frac{\Sigma^{(3)}_n}{\Sigma^{(2)}_n}(\tv-p_{n-1}).
\end{align}
Similarly, the part of the insider's orders $\Delta\theta^I_n$ which is expected by the rebalancer (and not the market makers) is given by
\begin{align}
\E[\Delta \theta_n^I | \sigma(\ta,y_1, \ldots, y_{n-1})]-\E[\Delta \theta_n^I | \sigma(y_1, \ldots, y_{n-1})]=\beta_{n}^I\frac{\Sigma^{(3)}_n}{\Sigma^{(1)}_n}(\ta-\theta^R_{n-1}-p_{n-1}).
\end{align}
This equality follows from the market makers not expecting the insider to trade at all.

\begin{figure}[!h]
\begin{center}
\caption{Plot of conditional expectations of the predictable parts of the rebalancer's trades (left is the market makers' estimate and right is the insider's estimate). These graphs are independent of $(\ta,\tv,w)$'s realization. The parameters are $\sigma^2_{\tv}=1$, $\sigma_w^2= 4$, $N=10$ and $\rho =0$. The variance of the trading target varies: $\sigma^2_{\ta}=0.48\; (-\cdot-\cdot-),\sigma^2_{\ta}=1\; (- - - ),   \sigma^2_{\ta}=3.7\;\text{(---------)}.$  }
$\begin{array}{cc}
\includegraphics[width=7cm, height=5cm]{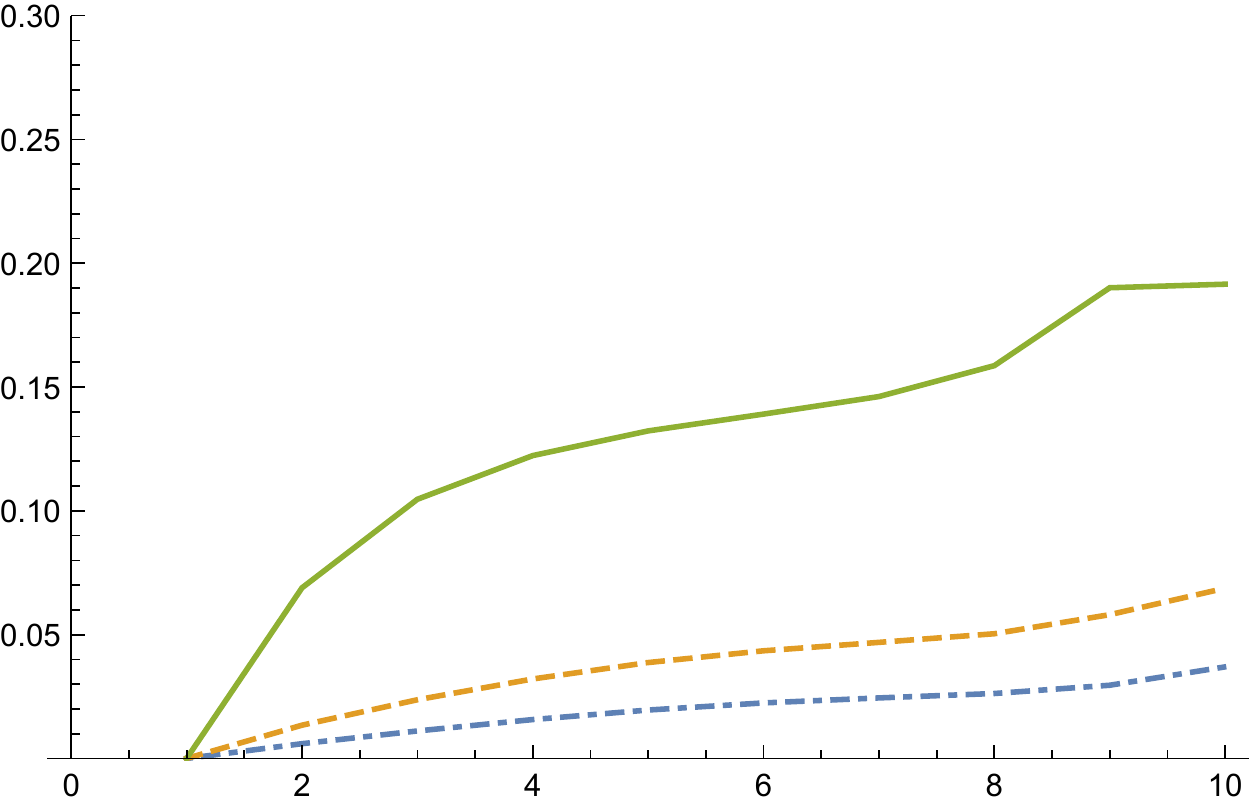} &
\includegraphics[width=7cm, height=5cm]{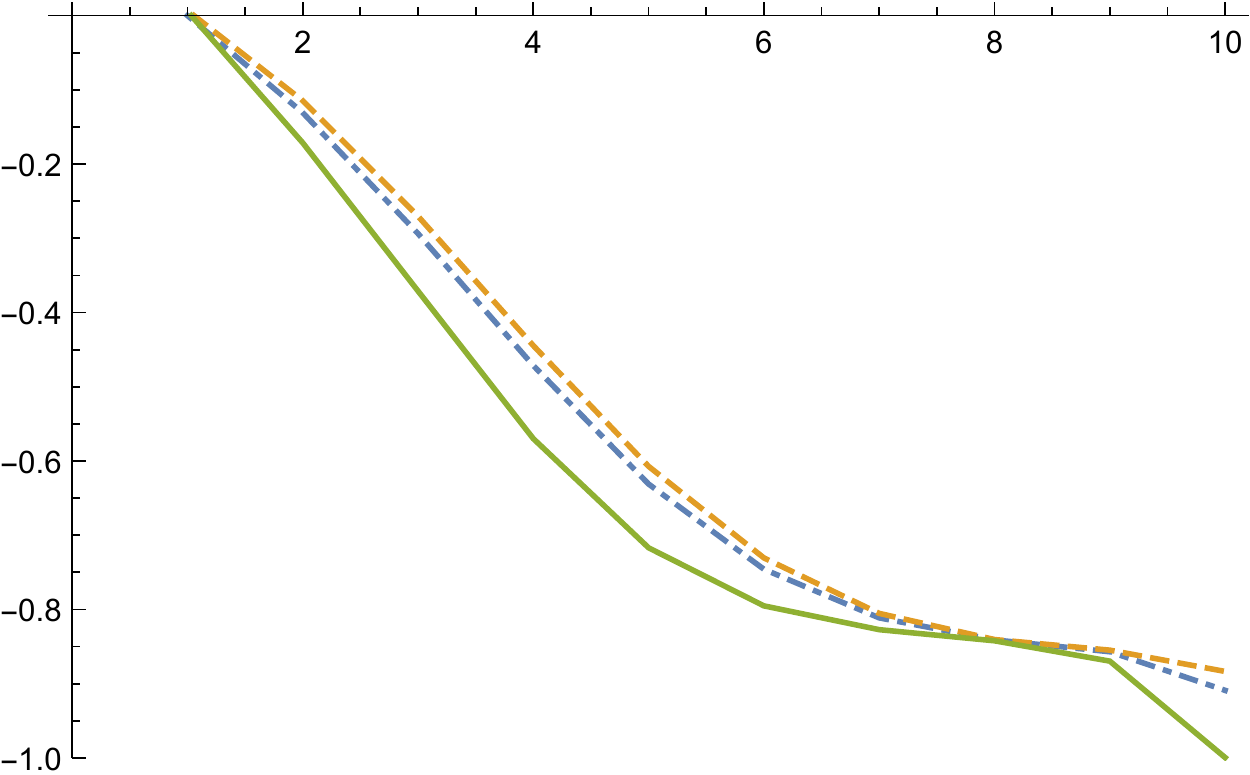}
\\
   \text{A: }\frac{\E\left[\E[\Delta \theta^R_n|\sigma(y_1,\cdots,y_{n-1})]|\sigma(\ta)\right]}{\E\left[\Delta \theta^R_n|\sigma(\ta)\right]}&\text{B: } \frac{\E[\Delta \theta^I_n|\sigma(\ta,y_1,\cdots,y_{n-1})]}{\Delta \theta^R_n - \E[ \Delta \theta^R_n | \sigma(y_1,\cdots,y_{n-1})]}\end{array}$
\label{fig:predict-Rorder-components}
\end{center}
\end{figure}

\begin{figure}[!h]
\begin{center}
\caption{Plot of corr($\Delta \theta^I_n,\Delta \theta^R_n)$ for $n=1,2,...,10$ (unconditional).  The parameters are $\sigma^2_{\tv}=1$, $\sigma_w^2= 4$, $N=10$, $\sigma^2_{\ta}=1$ (right only), and  $\rho =0$ (left only).}
$\begin{array}{cc}
\includegraphics[width=7cm, height=5cm]{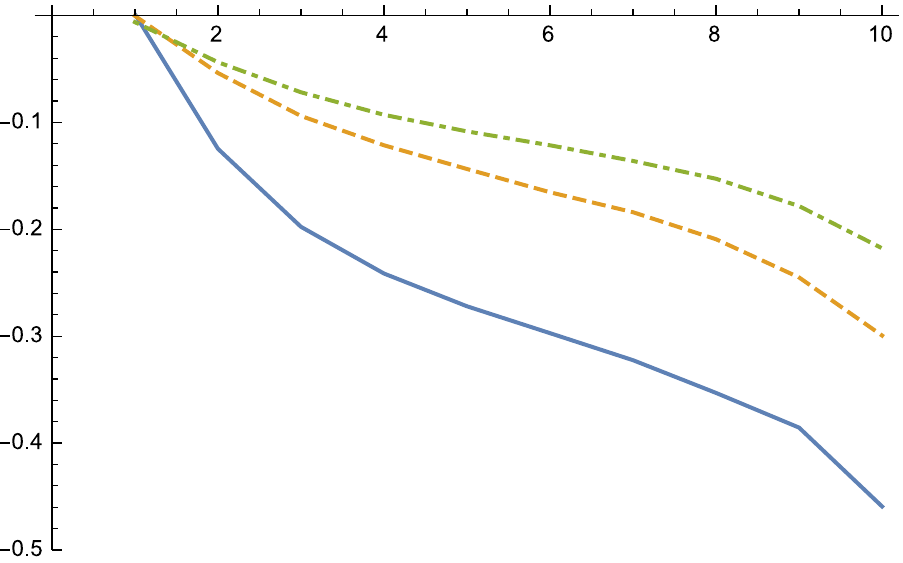} &
\includegraphics[width=7cm, height=5cm]{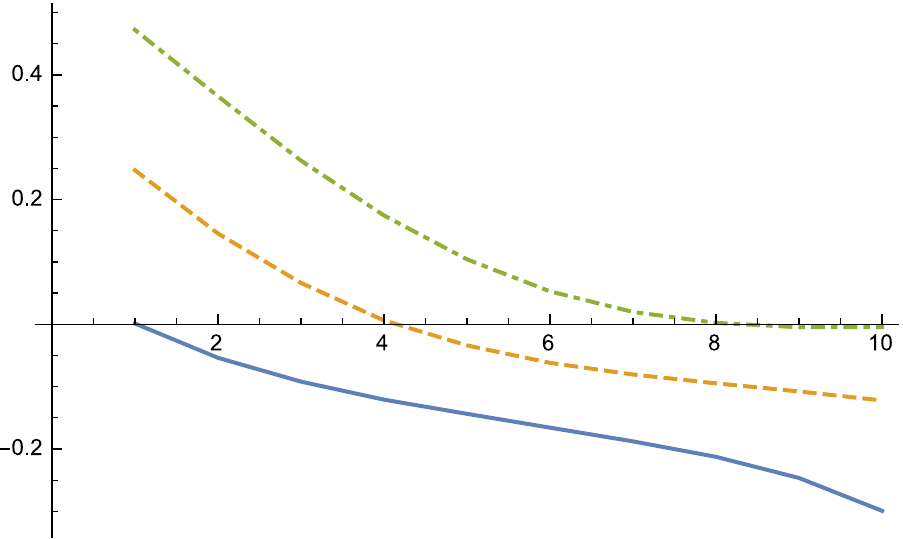}
\\
   \text{A: }\sigma^2_{\ta}=0.48\; (-\cdot-\cdot-),\;\sigma^2_{\ta}=1\; (- - - ), &\text{B: } \rho=0\; \;\text{(---------)},\;\rho=0.25\;(- - -),\\
  \sigma^2_{\ta}=3.7\;\text{(---------)}.   &\rho=0.47\;(-\cdot-\cdot-).\\
\end{array}$
\label{fig:order-corr}
\end{center}
\end{figure}

Figure \ref{fig:predict-Rorder-components}A measures the fraction of the rebalancer's order at time $n$ which he expects to trade over time in a way that is predictable for the market maker. We see that a sunshine component is present but is not particularly large (less than 5\% when $\sigma^2_{\ta}=1$).
Figure \ref{fig:predict-Rorder-components}B shows, however, that a portion of the unpredictable part of the rebalancer's orders (given the market makers' information) is predictably offset (given the rebalancer's information) by the insider's trades. This type of predictability is mutual beneficial for both the rebalancer and the insider. By trading in opposite directions, they provide symbiotically liquidity to each other with a reduced price impact. This is evidenced in Figure \ref{fig:order-corr}, which shows that the resulting conditional correlation between the insider's orders and the rebalancer's orders is negative later in the day.


\newpage
\section{Conclusion}

This paper has explored the equilibrium interactions between strategic dynamic informed trading, strategic dynamic portfolio rebalancing, price discovery, and liquidity in a multi--period Kyle (1985) market.  To the best of our knowledge, our paper is the first to investigate these issues with both long-lived information and dynamic rebalancing given a terminal trading target.  We find that uninformed trading has noteworthy effects on the market equilibrium.  Order flow becomes  autocorrelated and liquidity and price discovery dynamics change. We also show that, in equilibrium, optimal uninformed trading strategies involve learning about the informed trader's information and exploiting that information to reduce costs.  In addition, we find that, over time, there is an interesting negative correlation structure between the informed trader's and rebalancer's orders. Because the insider's and rebalancer's orders partially cancel each other, they can supply liquidity to each other symbioticallty with a reduced price impact.

There are many interesting possible extensions for future work.  One possible extension is to model trading in continuous-time. We could also consider other forms of portfolio rebalancing constraints. A third extension is to relax the assumption that all investors are risk-neutral. For this extension, it would be natural to consider exponential utilities with different  coefficients of absolute risk aversion. Finally, it would be interesting to extend the model to include multiple insiders and rebalancers.

\pagebreak

\appendix

\section{Proofs}\label{sec:proofs}

\subsection{Kalman filtering}\label{sec:Kalman}
\begin{lemma}\label{lem:Kalman} Consider the ``hat'' system \eqref{optimal_nheta_I}-\eqref{hatq} corresponding to arbitrary coefficients $(\beta^I_n,\beta^R_n,\alpha^R_n)_{n=1}^N$. Whenever \eqref{ass1lambda}-\eqref{ass1s} hold, we have
\begin{align}
\hat{p}_n &= \E[ \tv |\sigma(\hy_1,...,\hy_n)],\\
\hat{q}_n &= \E[ \tilde{a}-\hat{\theta}^R_n |\sigma(\hy_1,...,\hy_n)],
\end{align}
where $\hat{p}$ is defined by \eqref{hatp} and $\hat{q}$ is defined by \eqref{hatq}. Furthermore, the recursions for the variances and covariance \eqref{S1iterative}-\eqref{S3iterative} hold.
\end{lemma}
\proof
For $n=1,...,N$, we have the moment definitions in \eqref{S11}-\eqref{S33} where the starting values are given in (\ref{eq:starting-values}).
We then define the process $\hz^M_n$ as
\begin{equation}\begin{split}
\hz^M_n:=&\hy_n - (\alpha_n^R+\beta_n^R)\hq_{n-1}\\
=& \beta_n^I (\tv-\hp_{n-1}) +\beta_n^R (\ta-\hat{\theta}^R_{n-1}-\hq_{n-1})     +\Delta w_n.\label{hz}
\end{split}\end{equation}
These Gaussian variables $\hz^M_1,\hz^M_2,....,\hz^M_N$ are mutually  independent and satisfy $\sigma(\hz^M_1,\\...,\hz^M_n) = \sigma(\hy_1,...\hy_n)$. The projection theorem for Gaussian random variables produces
\begin{align*}
\Delta\hp_n=& \E[\tv|\sigma(\hz^M_1,...,\hz^M_n)]- \E[\tv|\sigma(\hz^M_1,...,\hz_{n-1})]\\
=& \frac{\E[\tv\, \hz^M_n]   }{\V[\hz^M_n]} \hz^M_n ,\\
\Delta \hq_n= &\E[\ta - \hat{\theta}^R_n|\sigma(\hz^M_1,...\hz^M_n)]-\E[\ta - \hat{\theta}^R_{n-1}|\sigma(\hz^M_1,...,\hz^M_{n-1})]\\
=&\E[\ta - \hat{\theta}^R_{n-1}|\sigma(\hz^M_1,...\hz^M_n)]-\E[\ta - \hat{\theta}^R_{n-1}|\sigma(\hz^M_1,...,\hz^M_{n-1})]  - \E[\Delta\hat{\theta}^R_{n}|\sigma(\hz^M_1,...,\hz^M_n)] \\
=& \frac{\E[(\ta - \hat{\theta}^R_{n-1}) \hz^M_n]   }{\V[\hz^M_n]} \hz^M_n - \E\big[ \beta_n^R(\ta - \hat{\theta}_{n-1}^R - \hq_{n-1}) +( \alpha_n^R+\beta_n^R) \hq_{n-1}  \big|\sigma(\hz^M_1,...,\hz^M_n)\big] \\
=&\frac{\E[(\ta - \hat{\theta}^R_{n-1}-\hq_{n-1}) \hz^M_n]   }{\V[\hz^M_n]} \hz^M_n - \beta_n^R \E[\ta - \hat{\theta}_{n-1}^R - \hq_{n-1}|\sigma(\hz^M_{n})] - (\alpha_n^R +\beta_n^R)\hq_{n-1} \\
=&(1-\beta_n^R) \frac{\E[(\ta - \hat{\theta}^R_{n-1}-\hq_{n-1}) \hz^M_n]   }{\V[\hz^M_n]} \hz^M_n - (\alpha_n^R+\beta_n^R) \hq_{n-1}.
\end{align*}
To proceed, we first need to compute
\begin{align*}
\V[\hz^M_n]=&\E\Big[ \Big(    \beta_n^I (\tv-\hp_{n-1}) +\beta_n^R (\ta-\hat{\theta}^R_{n-1}-\hq_{n-1})     +\Delta w_n         \Big)^2\Big]\\
=& (\beta_n^I)^2 \Sigma_{n-1}^{(2)}+ (\beta_n^R)^2 \Sigma_{n-1}^{(1)} + 2 \beta_n^I \beta_n^R \Sigma_{n-1}^{(3)}  + \sigma_w^2\Delta,\\
\E[\tv \hz^M_n]=&\E[(\tv -\hp_{n-1})\hz^M_n]\\
=&\E\left[(\tv -\hp_{n-1}) \Big(    \beta_n^I (\tv-\hp_{n-1}) +\beta_n^R (\ta-\hat{\theta}^R_{n-1}-\hq_{n-1})     +\Delta w_n         \Big)\right]\\
=& \beta_n^I \Sigma_{n-1}^{(2)} + \beta_n^R \Sigma_{n-1}^{(3)},\\
\E[(\ta - \hat{\theta}^R_{n-1}-\hq_{n-1}) \hz^M_n]=&\E\left[(\ta - \hat{\theta}^R_{n-1}-\hq_{n-1})  \Big(    \beta_n^I (\tv-\hp_{n-1}) +\beta_n^R (\ta-\hat{\theta}^R_{n-1}-\hq_{n-1})     +\Delta w_n         \Big)\right]\\
=&\beta_n^I \Sigma_{n-1}^{(3)} + \beta_n^R \Sigma_{n-1}^{(1)}.
\end{align*}
Combining these expressions and by matching coefficients with \eqref{hatp} and \eqref{hatq}, we find the lemma's statement equivalent to the restrictions \eqref{ass1lambda}-\eqref{ass1s}.
Based on these expressions, the recursion for $\Sigma_n^{(1)}$, $n=1,...,N$, in (\ref{S1iterative}) is
\begin{align*}
\Sigma^{(1)}_{n} :&= \V[\ta-\hat{\theta}_{n}^R-\hq_{n}]\nonumber\\
&=\V[\ta-\hat{\theta}_{n-1}^R-\hq_{n-1} - \Delta \hat{\theta}^R_n -\Delta \hq_n]\nonumber\\
&=\V[\ta-\hat{\theta}_{n-1}^R-\hq_{n-1} - \Delta \hat{\theta}^R_n -r_n \hy_n -s_n \hq_{n-1}]\nonumber\\
 &=\V\Big[\ta-\hat{\theta}_{n-1}^R-(1+s_n) \hq_{n-1} -(1+r_n)(\beta_n^R(\ta -\hat{\theta}_{n-1}^R) + \alpha_n^R \hq_{n-1}) \nonumber\\&\quad \quad - r_n\big(\beta_n^I(\tv -\hp_{n-1}) \big)-r_n\Delta w_n\Big],\nonumber\\
 &=\V\Big[\big(1-(1+r_n)\beta_n^R\big)(\ta-\hat{\theta}_{n-1}^R)-\big(1+s_n+(1+r_n)\alpha_n^R\big) \hq_{n-1} \nonumber\\&\quad \quad - r_n\beta_n^I(\tv -\hp_{n-1}) -r_n\Delta w_n\Big]\nonumber\\
  &=\V\Big[\big(1-(1+r_n)\beta_n^R\big)(\ta-\hat{\theta}_{n-1}^R-\hq_{n-1}) - r_n\beta_n^I(\tv -\hp_{n-1}) -r_n\Delta w_n\Big]\nonumber\\
  &= \big(1-(1+r_n)\beta_n^R\big)^2 \Sigma^{(1)}_{n-1} + (r_n\beta_n^I)^2 \Sigma^{(2)}_{n-1} + r_n^2\sigma_w^2\Delta - 2\big(1-(1+r_n)\beta_n^R\big)r_n\beta_n^I\Sigma^{(3)}_{n-1}\nonumber\\
  &=(1-\beta_n^R)\big( (1-\beta_n^R-r_n\beta_n^R)\Sigma_{n-1}^{(1)} - r_n \beta_n^I \Sigma_{n-1}^{(3)}  \big),
\end{align*}
where the last equality uses \eqref{ass1r}. The recursions for $\Sigma^{(2)}_{n}$ and $\Sigma^{(3)}_{n}$, $n=1,...,N$, in \eqref{S2iterative} and \eqref{S3iterative} are found similarly.

$\endproof$

\subsection{Insider's optimization problem}


We start with the following lemma which contains most of the calculations we will need later. 
We recall the insider's state processes $(X^{(1)}_n,X^{(2)}_n)$ are defined by \eqref{X}.

\begin{lemma}\label{Iinfo} Fix $\Delta \theta_n^R$ by \eqref{theta_L} and let the constants \eqref{all_constants} and associated terms \eqref{thm_constants} satisfy the pricing coefficient relations \eqref{ass1lambda}-\eqref{ass1s} and the variances and covariance recursions \eqref{S1iterative}-\eqref{S3iterative}. Let $\Delta \theta_n ^I \in \sigma(\tv,y_1,...,y_{n-1})$, $n=1,...,N$, be arbitrary for the insider.
We define the Gaussian random variables
\begin{align}
\hz^I_n:=&\hy_n - \Delta \hat{\theta}^I_n - (\alpha_n^R+\beta_n^R)\hq_{n-1}-\beta_n^R \tfrac{\Sigma^{(3)}_{n-1}}{\Sigma^{(2)}_{n-1}}(\tv-\hat{p}_{n-1}), \quad n=1,...,N.\label{hzI}
\end{align}
Then $\hz^I_k$ is independent of $(\ta,\hy_1,....,\hy_{k-1})$ for $k\le N$ and the following measurability properties are satisfied:
\begin{align}\label{Isame}
\hat{\theta}_n^R - \theta_n^R \in \sigma(\tv, y_1,...,y_n)=\sigma(\tv,\hat{y}_1,...,\hy_n)=\sigma(\tv, \hz^I_1,...\hz^I_n),\quad n=1,...,N.
\end{align}
Furthermore, for $n=1,...,N$, we have the Markovian dynamics
\begin{align}
\Delta X^{(1)}_{n}
& = -\lambda_n \Big( \Delta \theta_n^I + \beta^R_nX^{(2)}_{n-1}\Big)-\lambda_n  \hz^I_n, \quad X_0^{(1)} =\tv, \label{dX1}\\
\Delta X^{(2)}_n
& =  - r_n \Delta \theta^I_n -(1+r_n)\beta^R_{n}X^{(2)}_{n-1} -  \frac{\Sigma^{(3)}_{n}}{\Sigma^{(2)}_{n}}\lambda_n\hz^I_n,\quad  \quad X_0^{(2)} =\frac{\rho\sigma_{\ta}}{\sigma_{\tv}}\tv.\label{dX2}
\end{align}
Finally, for any constants $I_n^{(1,1)}, I_n^{(1,2)}$, and $I_n^{(2,2)}$, we have the conditional expectation
\begin{footnotesize}
\begin{align}
&\E\left[(\tv -p_n)\Delta\theta^I_{n}+ I_n^{(1,1)} \left(X_n^{(1)}\right)^2+ I_n^{(1,2)} X_n^{(1)}X_n^{(2)}+ I_n^{(2,2)} \left(X_n^{(2)}\right)^2\Big|\sigma(\tv,y_1,...,y_{n-1})\right]\nonumber\\
&= X^{(1)}_{n-1}\Delta\theta_n^I-(\Delta\theta_n^I)^2\lambda_n-\Delta\theta_n^I\lambda_n\beta^R_nX^{(2)}_{n-1} \nonumber\\
&+ I_n^{(1,1)}\Big(\left(X_{n-1}^{(1)}\right)^2 -2\lambda_n X_{n-1}^{(1)} \Big( \Delta \theta^I_n + \beta^R_nX^{(2)}_{n-1}\Big) +\lambda_n^2 \Big( \Delta \theta^I_n+ \beta^R_nX^{(2)}_{n-1}\Big)^2 + \lambda^2_n\V[\hz^I_n]\Big)\nonumber\\
&+ I_n^{(1,2)}\Big(X_{n-1}^{(1)}X_{n-1}^{(2)} -X_{n-1}^{(1)}\Big( r_n \Delta \theta^I_n +(1+r_n) \beta^R_n X^{(2)}_{n-1}\Big)- X_{n-1}^{(2)}\lambda_n \Big( \Delta \theta_n^I + \beta^R_nX^{(2)}_{n-1}\Big)\label{Iconditional1}
\\& + \lambda_n \Big( \Delta \theta_n^I + \beta^R_nX^{(2)}_{n-1}\Big)\Big( r_n \Delta \theta^I_n+(1+r_n) \beta^R_n X^{(2)}_{n-1}\Big) +\lambda_n^2\tfrac{\Sigma_{n}^{(3)}}{\Sigma_{n}^{(2)}}\V[\hz^I_n] \Big)\nonumber\\
&+ I_n^{(2,2)}\Big(\left(X_{n-1}^{(2)}\right)^2 - 2X_{n-1}^{(2)} \Big(r_n \Delta \theta^I_n +(1+r_n) \beta^R_nX^{(2)}_{n-1}\Big) +\Big( r_n\Delta \theta^I_n+ (1+r_n)\beta^R_nX^{(2)}_{n-1}\Big)^2 \nonumber\\&+ \lambda^2_n\left(\tfrac{\Sigma_n^{(3)}}{\Sigma_n^{(2)}}\right)^2\V[\hz^I_n]\Big),\nonumber
\end{align}
\end{footnotesize}
which is quadratic in $\Delta \theta^I_n$, and where the variance $\V[\hz^I_n]$  can be computed to be
\begin{align}\label{varianceLemI}
\V[\hz^I_n] = (\beta^R_n)^2\Big(\Sigma_{n-1}^{(1)} - \tfrac{\big(\Sigma_{n-1}^{(3)}\big)^2}{\Sigma_{n-1}^{(2)}}\Big)+ \sigma_w^2\Delta.
\end{align}

\end{lemma}
\proof
The joint normality claim follows by an induction argument. To see the independence claim we start by noticing
\begin{align*}
& \beta^R_n\Big( \ta -\hat{\theta}^R_{n-1}-\hat{q}_{n-1} - \E\left[\ta -\hat{\theta}^R_{n-1}-\hat{q}_{n-1} |\sigma(\tv,\hy_1,...,\hy_{n-1})\right]\Big)+ \Delta w_n\\&=\beta_n^R\Big( \ta -\hat{\theta}^R_{n-1}-\hat{q}_{n-1}- \tfrac{\Sigma_{n-1}^{(3)}}{\Sigma_{n-1}^{(3)}} (\tv-\hat{p}_{n-1})\Big)+\alpha_n^R\hat{q}_{n-1} -\alpha_n^R\hat{q}_{n-1} + \Delta w_n\\
&= \hy_n -\Delta \hat{\theta}^I_n -(\alpha_n^R+\beta_n^R) \hq_{n-1}-\beta_n^R \tfrac{\Sigma_{n-1}^{(3)}}{\Sigma_{n-1}^{(3)}} \big(\tv- \hat{p}_{n-1}\big),
\end{align*}
which is $\hz^I_n$ (see \ref{hzI}). To see the independence of the random variables \eqref{hzI} we let $k\le n-1$ be arbitrary.  Iterated expectations produce
the zero correlation property:
\begin{align*}
&\E[\hat{y}_k\hat{z}^I_n] = \E[\E[\hat{y}_k\hat{z}^I_n|\sigma(\ta,\hy_1,...,\hy_k)]]= \E[\hat{y}_k\E[\hat{z}^I_n|\sigma(\ta,\hy_1,...,\hy_k)]]=0.
\end{align*}
The independence then follows from the joint normality.

Next, we observe that the last equality in \eqref{Isame} follows directly from \eqref{hzI}. We proceed by induction and observe
\begin{displaymath}\begin{split}
&\sigma(\tv,y_1)=\sigma(\tv,\beta_1^R\ta + \Delta w_1) = \sigma(\tv,\hy_1),\\
&\hat{\theta}_1^R - \theta_1^R=0,
\end{split}\end{displaymath}
which follows from $\hat{\theta}_1^I, \theta_1^I\in\sigma(\tv)$. Suppose that \eqref{Isame} holds for $n$. Then,
\begin{displaymath}\begin{split}
\hat{\theta}_{n+1}^R - \theta_{n+1}^R &= (1-\beta_{n+1}^R)(\hat{\theta}_{n}^R - \theta_{n}^R) + \alpha_{n+1}^R (\hq_n-q_n) \in \sigma(\tv,y_1,...,y_{n}),\\
\sigma(\tv,\hy_1,...,\hy_{n+1})&=\sigma(\tv,y_1,...,y_n,\hy_{n+1})\\
&=\sigma(\tv,y_1,...,y_n,y_{n+1}+ \Delta\hat{\theta}_{n+1}^I - \Delta\theta_{n+1}^I + \Delta\hat{\theta}_{n+1}^R - \Delta\theta_{n+1}^R )\\
&=\sigma(\tv,y_1,...,y_{n+1}),
\end{split}\end{displaymath}
which proves \eqref{Isame}.  The dynamics \eqref{dX1} can be seen as follows
\begin{align*}
\Delta X^{(1)}_{n}
&=  -\Delta p_n\\
& = -\lambda_n \Big( \Delta \theta^I_{n} + \beta^R_{n}(\ta - \theta^R_{n-1}) + \alpha^R_{n} q_{n-1} +  \Delta w_n\Big) - \mu_n q_{n-1}\\
& = -\lambda_n \Big( \Delta \theta^I_{n} + \beta^R_{n}(\ta - \theta^R_{n-1}) + \alpha^R_{n} q_{n-1} +  \hy_n - \Delta\hat{\theta}^I_n-\Delta\hat{\theta}^R_n\Big) +\lambda_n(\alpha^R_n+\beta^R_n) q_{n-1}\\
& = -\lambda_n \Big( \Delta \theta^I_{n} + \beta^R_{n}(\hat{\theta}^R_{n-1} - \theta^R_{n-1}) + \hz^I_n  +\beta_n^R(\hq_{n-1}-q_{n-1})+\beta^R_n\tfrac{\Sigma^{(3)}_{n-1}}{\Sigma^{(2)}_{n-1}} \big(\tv- \hat{p}_{n-1}\big)
\Big) \\
& = -\lambda_n \Big( \Delta \theta_n^I + \beta^R_nX^{(2)}_{n-1}+  \hz^I_n\Big),
\end{align*}
The dynamics \eqref{dX2} are found similarly using expressions  \eqref{ass1lambda}-\eqref{ass1r} and \eqref{S2iterative}-\eqref{S3iterative}.

The expression for the variance \eqref{varianceLemI} is found as follows:
\begin{align*}
\V[\hz^I_n] &= \V\left[ \beta^R_n\Big( \ta -\hat{\theta}^R_{n-1}-\hat{q}_{n-1} - \E\left[\ta -\hat{\theta}^R_{n-1}-\hat{q}_{n-1} |\sigma(\tv,\hy_1,...,\hy_{n-1})\right]\Big)+ \Delta w_n\right]\\
&=\V\left[ \beta^R_n\Big( \ta -\hat{\theta}^R_{n-1}-\hat{q}_{n-1} - \tfrac{\Sigma_{n-1}^{(3)}}{\Sigma_{n-1}^{(2)}}(\tv-\hat{p}_{n-1})\Big)\right]+ \sigma_w^2\Delta \\
&= (\beta^R_n)^2\Big(\Sigma_{n-1}^{(1)} - \tfrac{\big(\Sigma_{n-1}^{(3)}\big)^2}{\Sigma_{n-1}^{(2)}}\Big)+ \sigma_w^2\Delta.
\end{align*}

To compute the conditional expectation  \eqref{Iconditional1}, we compute the four individual terms. The first term in \eqref{Iconditional1} equals
\begin{align*}
&\E[(\tv -p_n)\Delta\theta_n^I|\sigma(\tv,y_1,...,y_{n-1})]\\
&=  (\tv -p_{n-1})\Delta\theta^I_n-\Delta\theta^I_n\E[\Delta p_n|\sigma(\tv,y_1,...,y_{n-1})]\\
&= X^{(1)}_{n-1}\Delta\theta_n^I-\Delta\theta^I_n\lambda_n\E[ \Delta\theta^I_n+\beta^R_n(\ta-\theta^R_{n-1}-q_{n-1})|\sigma(\tv,y_1,...,y_{n-1})]\\
&= X^{(1)}_{n-1}\Delta\theta_n^I-(\Delta\theta^I_n)^2\lambda_n\\&\quad -\Delta\theta_n^I\lambda_n\beta^R_n\Big(\hat{\theta}^R_{n-1}-\theta^R_{n-1}+\hat{q}_{n-1}-q_{n-1}+\E[\ta-\hat{\theta}^R_{n-1}-\hat{q}_{n-1}|\sigma(\tv,y_1,...,y_{n-1})]\Big)\\
&=X^{(1)}_{n-1}\Delta\theta_n^I-(\Delta\theta^I_{n})^2\lambda_n-\Delta\theta^I_n\lambda_n\beta^R_n\Big(\hat{\theta}^R_{n-1}-\theta^R_{n-1}+\hat{q}_{n-1}-q_{n-1}+\frac{\Sigma^{(3)}_{n-1}}{\Sigma^{(2)}_{n-1}}(\tv-\hat{p}_{n-1})\Big)\\
&= X^{(1)}_{n-1}\Delta\theta_n^I-(\Delta\theta_n^I)^2\lambda_n-\Delta\theta_n^I\lambda_n\beta^R_nX^{(2)}_{n-1}.
\end{align*}
The second term in \eqref{Iconditional1} is
\begin{align*}
&\E[\left(X_n^{(1)}\right)^2|\sigma(\tv,y_1,...,y_{n-1})]\\
&=\left(X_{n-1}^{(1)}\right)^2 + 2X_{n-1}^{(1)}\E[\Delta X^{(1)}_n|\sigma(\tv,y_1,...,y_{n-1})]+\E[\left(\Delta X^{(1)}_n\right)^2|\sigma(\tv,y_1,...,y_{n-1})]\\
&=\left(X_{n-1}^{(1)}\right)^2 -2\lambda_n X_{n-1}^{(1)} \Big( \Delta \theta^I_n + \beta^R_nX^{(2)}_{n-1}\Big) +\lambda_n^2 \Big( \Delta \theta^I_n+ \beta^R_nX^{(2)}_{n-1}\Big)^2 + \lambda^2_n\V[\hz^I_n].
\end{align*}
The third term in \eqref{Iconditional1} is
\begin{align*}
&\E[X_n^{(1)}X_n^{(2)}|\sigma(\tv,y_1,...,y_{n-1})]\\
&=X_{n-1}^{(1)}X_{n-1}^{(2)} + X_{n-1}^{(1)}\E[\Delta X_{n}^{(2)}|\sigma(\tv,y_1,...,y_{n-1})]+ X_{n-1}^{(2)}\E[\Delta X_{n}^{(1)}|\sigma(\tv,y_1,...,y_{n-1})] \\&+ \E[\Delta X_{n}^{(1)}\Delta X_{n}^{(2)}|\sigma(\tv,y_1,...,y_{n-1})]\\
&=X_{n-1}^{(1)}X_{n-1}^{(2)} -X_{n-1}^{(1)}\Big( r_n \Delta \theta^I_n +(1+r_n) \beta^R_n X^{(2)}_{n-1}\Big)- X_{n-1}^{(2)}\lambda_n \Big( \Delta \theta_n^I + \beta^R_nX^{(2)}_{n-1}\Big)\\& + \lambda_n \Big( \Delta \theta_n^I + \beta^R_nX^{(2)}_{n-1}\Big)\Big( r_n \Delta \theta^I_n+(1+r_n) \beta^R_n X^{(2)}_{n-1}\Big) +\lambda_n^2\tfrac{\Sigma_{n}^{(3)}}{\Sigma_{n}^{(2)}}\V[\hz^I_n].
\end{align*}
Finally, the last term in \eqref{Iconditional1} is
\begin{align*}
&\E[\left(X_n^{(2)}\right)^2|\sigma(\tv,y_1,...,y_{n-1})]\\
&=\left(X_{n-1}^{(2)}\right)^2 + 2X_{n-1}^{(2)}\E[\Delta X^{(2)}_n|\sigma(\tv,y_1,...,y_{n-1})]+\E[\left(\Delta X^{(2)}_n\right)^2|\sigma(\tv,y_1,...,y_{n-1})]\\
&=\left(X_{n-1}^{(2)}\right)^2 - 2X_{n-1}^{(2)} \Big(r_n \Delta \theta^I_n +(1+r_n) \beta^R_nX^{(2)}_{n-1}\Big) \\&+\Big( r_n\Delta \theta^I_n+ (1+r_n)\beta^R_nX^{(2)}_{n-1}\Big)^2 + \lambda^2_n\left(\tfrac{\Sigma_n^{(3)}}{\Sigma_n^{(2)}}\right)^2\V[\hz^I_n].
\end{align*}
$\endproof$

\begin{theorem}\label{Ithm} Fix $\Delta \theta^R_n$ by \eqref{theta_L} and let the constants \eqref{all_constants} and associate terms \eqref{thm_constants} satisfy the pricing coefficient relations \eqref{ass1lambda}-\eqref{ass1s}, the variances and covariance recursions \eqref{S1iterative}-\eqref{S3iterative}, the value function coefficient recursions \eqref{I^{(1,1)}_n}-\eqref{I^{(2,2)}_n} and the second-order-condition \eqref{Isecond}. Then the insider's value function has the quadratic form \eqref{Ivalue-body}
where $X_n^{(1)}$ and $X_n^{(2)}$ are defined in \eqref{X} and $\Delta p_n$ is defined by \eqref{eq_p}. Furthermore, the insider's optimal trading strategy is given by \eqref{Iopt}.

\end{theorem}
\proof We prove the theorem by the backward induction. Suppose that \eqref{Ivalue-body} 
holds for $n+1$. The insider's value function in the $n$'th iteration then becomes
\begin{equation}\begin{split}\label{Imax}
&\max_{\stackrel{\Delta\theta_k^I\in \sigma(\tv,y_1,...,y_{k-1})}{n\leq k \leq N}} \E \Big[ \sum_{k=n}^N (\tv-p_k)\Delta \theta^I_k \Big|\sigma( \tv,y_1,...,y_{n-1} )\Big]\\
&=\max_{   \Delta\theta_n^I \in \sigma(\tv,y_1,...,y_{n-1})} \E\Big[ (\tv-p_n)\Delta\theta_n^I+ I_n^{(0)} + \sum_{1\leq i \leq j \leq 2} I_n^{(i,j)} X_n^{(i)}X_n^{(j)}\Big| \sigma(\tv,y_1,...,y_{n-1} )\Big].\\
\end{split}\end{equation}
Because \eqref{Isecond} holds, Lemma \ref{Iinfo} shows that the coefficient in front of $(\Delta \theta_n^R)^2$ appearing in \eqref{Imax} is strictly negative. Consequently, the first-order condition is sufficient for optimality and the maximizer is \eqref{Iopt}. The value function coefficient recursions \eqref{I^{(1,1)}_n}-\eqref{I^{(2,2)}_n} are obtained by inserting the optimizer \eqref{Iopt} into \eqref{Imax}.

$\endproof$

\subsection{Rebalancer's optimization problem}

In the following analogue of Lemma \ref{Iinfo} we recall that the rebalancer's state variables $(Y^{(1)}_n,Y^{(2)}_n,Y^{(3)}_n)$ are defined in \eqref{Y}.


\begin{lemma}\label{Linfo} We define $\Delta \theta_n^I$ by \eqref{theta_I} and let the constants \eqref{all_constants} and associate terms \eqref{thm_constants} satisfy the pricing coefficient relations \eqref{ass1lambda}-\eqref{ass1s} and  the variances and covariance recursions \eqref{S1iterative}-\eqref{S3iterative}. Let $\Delta \theta_n ^R \in \sigma(\ta,y_1,...,y_{n-1})$, $n=1,...,N$ be arbitrary for the rebalancer. We define the Gaussian random variables
\begin{align}
\hz^R_n:=\hy_n - \Delta \hat{\theta}^R_n-\beta_n^I \tfrac{\Sigma^{(3)}_{n-1}}{\Sigma^{(1)}_{n-1}}(\ta-\hat{\theta}^R_{n-1}-\hat{q}_{n-1}), \quad n=1,...,N.\label{hzR}
\end{align}
Then $\hz^R_k$ is independent of $(\tv,\hy_1,....,\hy_{k-1})$ for $k\le N$ and the following measurability properties are satisfied
\begin{align}
&\sigma(\ta, y_1,...,y_k)=\sigma(\ta,\hat{y}_1,...,\hy_k)=\sigma(\ta, \hz^R_1,...,\hz^R_k) 
\label{Lsame}.
\end{align}
Furthermore, for $n=1,...,N$, we have the Markovian dynamics
\begin{align}
\Delta Y^{(2)}_{n}
& = -\lambda_n \Big( \Delta \theta_n^R + \beta^I_nY^{(2)}_{n-1}-(\alpha^R_n+\beta_n^R)Y^{(3)}_{n-1}\Big) - r_n\tfrac{\Sigma^{(3)}_n}{\Sigma^{(1)}_n} \hz^R_n, \quad Y_0^{(2)} =\frac{\sigma_{\tv}\rho}{\sigma_{\ta}}\ta, \label{dY2}\\
\Delta Y^{(3)}_n
& =  r_n\Big( \Delta \theta^R_n + \beta^I_{n}Y^{(2)}_{n-1} \Big) -(1+r_n)( \alpha_n^R+\beta_n^R)Y^{(3)}_{n-1}+r_n\hz^R_n,  \quad Y_0^{(3)} =0. \label{dY3}
\end{align}
For constants $L^{(1,1)}_n,L^{(1,2)}_n,L^{(1,3)}_n,L^{(2,2)}_n,L^{(2,3)}_n$, and $L^{(3,3)}_n$ we have the conditional expectation
\begin{footnotesize}
\begin{align}
&\E[-(\ta - \theta^R_{n-1}) \Delta p_n + \sum_{1\leq i \leq j \leq 3} L_n^{(i,j)} Y_n^{(i)}Y_n^{(j)}|\sigma(\ta,y_1,...,y_{n-1})] \nonumber\\&=
 -Y_{n-1}^{(1)}\Big( \lambda_n(\Delta \theta_n^R + \beta^I_nY^{(2)}_{n-1}) + \mu_n Y^{(3)}_{n-1}\Big)\nonumber\\
 &+ L^{(1,1)}_n\Big((Y^{(1)}_{n-1} - \Delta \theta^R_n)^2\Big)\nonumber\\
 &+ L^{(1,2)}_n(Y^{(1)}_{n-1} - \Delta \theta^R_n)\big(Y^{(2)}_{n-1}-\lambda_n \big( \Delta \theta_n^R + \beta^I_nY^{(2)}_{n-1}-(\alpha^R_n+\beta_n^R)Y^{(3)}_{n-1}\big)\big)\nonumber\\
 &+ L^{(1,3)}_n(Y^{(1)}_{n-1} - \Delta \theta^R_n)\big(Y^{(3)}_{n-1}+ r_n\big( \Delta \theta^R_n + \beta^I_{n}Y^{(2)}_{n-1}\big) -(1+r_n)(\alpha_n^R+\beta_n^R)Y^{(3)}_{n-1}\big)\nonumber\\
 &+ L^{(2,2)}_n\Big((Y^{(2)}_{n-1})^2 -2Y^{(2)}_{n-1}\lambda_n \Big( \Delta \theta_n^R + \beta^I_nY^{(2)}_{n-1}-(\alpha^R_n+\beta_n^R)Y^{(3)}_{n-1}\Big)\nonumber\\&\quad + \lambda_n^2 \Big( \Delta \theta_n^R + \beta^I_nY^{(2)}_{n-1}-(\alpha^R_n+\beta_n^R)Y^{(3)}_{n-1}\Big)^2 +r_n^2\left(\tfrac{\Sigma^{(3)}_n}{\Sigma^{(1)}_n}\right)^2 \V[\hz^R_n]\Big) \label{Lconditional1}\\
&+ L^{(2,3)}_n\Big(Y^{(2)}_{n-1}Y^{(3)}_{n-1} +Y^{(2)}_{n-1}
\Big(r_n\big( \Delta \theta^R_n + \beta^I_{n}Y^{(2)}_{n-1}\big) -(1+r_n)(\alpha_n^R+\beta_n^R)Y^{(3)}_{n-1}\Big) \nonumber\\
&\quad -Y^{(3)}_{n-1}\lambda_n \Big( \Delta \theta_n^R + \beta^I_nY^{(2)}_{n-1}-(\alpha^R_n+\beta_n^R)Y^{(3)}_{n-1}\Big)  -r_n^2\tfrac{\Sigma^{(3)}_n}{\Sigma^{(1)}_n} \V[\hz^R_n]\nonumber
\\
&\quad -\lambda_n \Big( \Delta \theta_n^R + \beta^I_nY^{(2)}_{n-1}-(\alpha^R_n+\beta_n^R)Y^{(3)}_{n-1}\Big)\Big(r_n\big( \Delta \theta^R_n + \beta^I_{n}Y^{(2)}_{n-1}\big) -(1+r_n)(\alpha_n^R+\beta_n^R)Y^{(3)}_{n-1}\Big)
\Big)\nonumber\\
&+ L^{(3,3)}_n\Big((Y^{(3)}_{n-1})^2 +2Y^{(3)}_{n-1}\Big(r_n\big( \Delta \theta^R_n + \beta^I_{n}Y^{(2)}_{n-1}\big) -(1+r_n)(\alpha_n^R+\beta_n^R)Y^{(3)}_{n-1}\Big)\nonumber\\
 &\quad+ \Big(r_n\big( \Delta \theta^R_n + \beta^I_{n}Y^{(2)}_{n-1}\big) -(1+r_n)(\alpha_n^R+\beta_n^R)Y^{(3)}_{n-1}\Big)^2 + r_n^2 \V[\hz_n^R]\Big),\nonumber
\end{align}
\end{footnotesize}
which is quadratic in $\Delta \theta^R_n$, and where the variance $\V[\hz_n^R]$ is given by
\begin{align}\label{varianceLemR}
\V[\hz^R_n] = (\beta^I_n)^2\Big(\Sigma_{n-1}^{(2)} - \tfrac{\big(\Sigma_{n-1}^{(3)}\big)^2}{\Sigma_{n-1}^{(1)}}\Big)+ \sigma_w^2\Delta.
\end{align}

\end{lemma}

\proof The proof is similar to the proof of Lemma \ref{Iinfo} and is therefore omitted.

$\endproof$

\begin{theorem}\label{Lthm} Fix $\Delta \theta^I_n$ by \eqref{theta_I} and let the constants \eqref{all_constants} and associated terms \eqref{thm_constants} satisfy the pricing coefficient relations \eqref{ass1lambda}-\eqref{ass1s}, the variances and covariance recursions \eqref{S1iterative}-\eqref{S3iterative}, the value function coefficient recursions \eqref{L11}-\eqref{L33} and the second-order-condition \eqref{Lsecond}. Then for $n=0,1,...,N-1$ the rebalancer's value function has the quadratic form \eqref{Lvalue-body}
where $(Y_n^{(1)},Y_n^{(3)},Y_n^{(3)})$ are defined by \eqref{Y}  and $\Delta p_n$ is defined by \eqref{eq_p}. Furthermore, the rebalancer's optimal trading strategy is given by \eqref{Lopt}.
\end{theorem}

\proof The proof is similar to the proof of Theorem \ref{Ithm} and is therefore omitted.

$\endproof$
\subsection{Remaining proof}

\proof[Proof of Theorem \ref{thm:main}] Part (iii) of Definition \ref{def:BayNash} holds from Lemma \ref{lem:Kalman}. Parts (i)-(ii) of Definition \ref{def:BayNash} hold from Theorem \ref{Ithm} and Theorem \ref{Lthm} as soon as we show that the optimizers  \eqref{Iopt} and \eqref{Lopt} agree with \eqref{optimal_nheta_I} and \eqref{optimal_nheta_L}. This, however, follows from the equilibrium conditions \eqref{eq:betaI} and \eqref{eq:betaRalpha}.

$\endproof$

\section{Value function coefficients}\label{sec:recursion}
The recursion for the insider's value function coefficients is given by
\begin{footnotesize}
\begin{align}
I_{n-1}^{(1,1)} &= \frac{-1 + r_n (2 I^{(1,2)}_n - (I^{(1,2)}_n)^2 r_n +
    4 I^{(1,1)}_n I^{(2,2)}_n r_n)}{4 (I^{(2,2)}_n r_n^2 + \lambda_n (-1 + I^{(1,2)}_n r_n +
      I^{(1,1)}_n \lambda_n))},\label{I^{(1,1)}_n}\\
I_{n-1}^{(1,2)} &= -\frac{(-1 + I^{(1,2)}_n r_n) (I^{(1,2)}_n (-1 + \beta^R_n) + \beta^R_n) \lambda_n +
  2 I^{(2,2)}_n r_n (-1 + \beta^R_n + r_n \beta^R_n -
     2 I^{(1,1)}_n (-1 + \beta^R_n) \lambda_n)}{2 (I^{(2,2)}_n r_n^2 + \lambda_n (-1 + I^{(1,2)}_n r_n +
      I^{(1,1)}_n \lambda_n))},\label{I^{(1,2)}_n}\\
I_{n-1}^{(2,2)} &=    \lambda_n\frac{ -(I^{(1,2)}_n (-1 + \beta^R_n) + \beta^R_n)^2 \lambda_n -
   4 I^{(2,2)}_n (-1 + \beta^R_n) (-1 +
      I^{(1,1)}_n \lambda_n + \beta^R_n (1 + r_n - I^{(1,1)}_n \lambda_n))}{4 (I^{(2,2)}_n r_n^2 + \lambda_n (-1 + I^{(1,2)}_n r_n +
      I^{(1,1)}_n \lambda_n))}.
\label{I^{(2,2)}_n}
\end{align}
\end{footnotesize}
The recursion for the rebalancer's value function coefficients is given by
\begin{footnotesize}
\begin{align}
\begin{split}\label{L11}
L_{n-1}^{(1,1)}
&=-\Big((L^{(1,3)}_n)^2 r_n^2 -
  2 (1 + L^{(1,2)}_n) L^{(1,3)}_n r_n \lambda_n + (1 + L^{(1,2)}_n)^2 \lambda_n^2
 \\&+
  4 L^{(1,1)}_n (-L^{(3,3)}_n r_n^2 + \lambda_n + L^{(2,3)}_n r_n \lambda_n - L^{(2,2)}_n \lambda_n^2)\Big)\Big/
  \\  &
 4 \big(L^{(1,1)}_n - L^{(1,3)}_n r_n + L^{(3,3)}_n r_n^2 + \lambda_n (L^{(1,2)}_n - L^{(2,3)}_n r_n + L^{(2,2)}_n \lambda_n)\big),
\end{split}
\\
 \begin{split}\label{L12}
L_{n-1}^{(1,2)}
&=
-\Big((L^{(1,3)}_n r_n - \lambda_n) (L^{(2,3)}_n r_n + L^{(1,3)}_n r_n \beta^I_n - 2 L^{(2,2)}_n \lambda_n) +
     (L^{(1,2)}_n)^2 \lambda_n (-1 + \beta^I_n \lambda_n)
       \\&
       +L^{(1,2)}_n (r_n (L^{(1,3)}_n - 2 L^{(3,3)}_n r_n) + \lambda_n +
        r_n (L^{(2,3)}_n - 2 L^{(1,3)}_n \beta^I_n) \lambda_n + \beta^I_n \lambda_n^2)
        \\&
        +2 L^{(1,1)}_n (-r_n (L^{(2,3)}_n + 2 L^{(3,3)}_n r_n \beta^I_n) + (2 L^{(2,2)}_n + \beta^I_n + 2 L^{(2,3)}_n r_n \beta^I_n) \lambda_n \\&- 2 L^{(2,2)}_n \beta^I_n \lambda_n^2)\Big)\Big/2 \big(L^{(1,1)}_n - L^{(1,3)}_n r_n +
       L^{(3,3)}_n r_n^2 + \lambda_n (L^{(1,2)}_n - L^{(2,3)}_n r_n + L^{(2,2)}_n \lambda_n)\big),
\end{split}
\end{align}

\begin{align}
L_{n-1}^{(1,3)}
&=
\Big[(L^{(1,3)}_n)^2 r_n (-1 + \alpha^R_n + r_n \alpha^R_n + \beta^R_n + r_n \beta^R_n) \nonumber
\\&
+ (1 + L^{(1,2)}_n) \lambda_n \Big(-2 L^{(3,3)}_n r_n (-1 + \alpha^R_n + \beta^R_n) - L^{(2,3)}_n \lambda_n \nonumber \\&+ (L^{(1,2)}_n + L^{(2,3)}_n) (\alpha^R_n + \beta^R_n) \lambda_n\Big) +
   2 L^{(1,1)}_n \big(-2 L^{(3,3)}_n r_n (-1 + \alpha^R_n + r_n \alpha^R_n + \beta^R_n +
         r_n \beta^R_n)\nonumber\\& -
      L^{(2,3)}_n \lambda_n + (\alpha^R_n + \beta^R_n) \lambda_n (1 + L^{(2,3)}_n +
         2 L^{(2,3)}_n r_n - 2 L^{(2,2)}_n \lambda_n)\big) \label{L13}
         \\&+
   L^{(1,3)}_n \lambda_n \Big(-1 + \alpha^R_n + \beta^R_n +
      L^{(2,3)}_n r_n (-1 + \alpha^R_n + \beta^R_n) \nonumber\\&-
      L^{(1,2)}_n (-1 + \alpha^R_n + 2 r_n \alpha^R_n + \beta^R_n +
         2 r_n \beta^R_n) +
      2 L^{(2,2)}_n \lambda_n - (\alpha^R_n + \beta^R_n) (r_n +
         2 L^{(2,2)}_n \lambda_n)\Big)\Big]\Big/\nonumber\\&2 (L^{(1,1)}_n - L^{(1,3)}_n r_n +
     L^{(3,3)}_n r_n^2 + \lambda_n (L^{(1,2)}_n - L^{(2,3)}_n r_n + L^{(2,2)}_n \lambda_n))\nonumber,\\
\begin{split}
L_{n-1}^{(2,2)} \label{L22}
&=
-\Big[(L^{(1,2)}_n)^2 (-1 + \beta^I_n \lambda_n)^2 -
     2 L^{(1,2)}_n r_n \big(L^{(2,3)}_n -
        L^{(2,3)}_n \beta^I_n \lambda_n \\&+ \beta^I_n (-L^{(1,3)}_n + 2 L^{(3,3)}_n r_n + L^{(1,3)}_n \beta^I_n \lambda_n)\big) +
     r_n \Big(\big((L^{(2,3)}_n)^2 - 4 L^{(2,2)}_n L^{(3,3)}_n\big) r_n \\&+ (L^{(1,3)}_n)^2 r_n (\beta^I_n)^2 + L^{(1,3)}_n (4 L^{(2,2)}_n + 2 L^{(2,3)}_n r_n \beta^I_n - 4 L^{(2,2)}_n \beta^I_n \lambda_n)\Big) \\&-
     4 L^{(1,1)}_n \big(L^{(2,2)}_n (-1 + \beta^I_n \lambda_n)^2 +
        r_n \beta^I_n (L^{(2,3)}_n + L^{(3,3)}_n r_n \beta^I_n -
           L^{(2,3)}_n \beta^I_n \lambda_n)\big)\Big]\Big/
           \\&4 (L^{(1,1)}_n - L^{(1,3)}_n r_n +
       L^{(3,3)}_n r_n^2 + \lambda_n (L^{(1,2)}_n - L^{(2,3)}_n r_n + L^{(2,2)}_n \lambda_n)),
\end{split}
       \\
L_{n-1}^{(2,3)}
&=
\Big[\big(L^{(1,3)}_n r_n (L^{(2,3)}_n + L^{(1,3)}_n \beta^I_n) -
      2 L^{(1,1)}_n (L^{(2,3)}_n + 2 L^{(3,3)}_n r_n \beta^I_n)\big) (-1 + \alpha^R_n +
      r_n \alpha^R_n + \beta^R_n +
      r_n \beta^R_n)
\nonumber      \\ &
      + \Big((L^{(2,3)}_n)^2 r_n (-1 + \alpha^R_n + \beta^R_n) +
      2 L^{(1,1)}_n L^{(2,3)}_n \beta^I_n (-1 + \alpha^R_n + 2 r_n \alpha^R_n + \beta^R_n +
         2 r_n \beta^R_n) \nonumber\\&+
      4 L^{(2,2)}_n (-L^{(3,3)}_n r_n (-1 + \alpha^R_n + \beta^R_n) +
         L^{(1,1)}_n (\alpha^R_n + \beta^R_n))
      \nonumber   \\&+
      L^{(1,3)}_n (L^{(2,3)}_n r_n \beta^I_n (-1 + \alpha^R_n + \beta^R_n) -
         2 L^{(2,2)}_n (1 + (-1 + r_n) \alpha^R_n + (-1 +
               r_n) \beta^R_n))\Big) \lambda_n
\nonumber               \\ &
               - 2 L^{(2,2)}_n \beta^I_n (L^{(1,3)}_n (-1 + \alpha^R_n + \beta^R_n) +
      2 L^{(1,1)}_n (\alpha^R_n + \beta^R_n)) \lambda_n^2 \nonumber\\&+
   (L^{(1,2)}_n)^2 (\alpha^R_n + \beta^R_n) \lambda_n (-1 + \beta^I_n \lambda_n)
   +
   L^{(1,2)}_n (L^{(2,3)}_n \lambda_n \label{L23} \\&-
      2 L^{(3,3)}_n r_n (-1 + \alpha^R_n + r_n \alpha^R_n + \beta^R_n +
         r_n \beta^R_n + \beta^I_n (-1 + \alpha^R_n + \beta^R_n) \lambda_n)\nonumber \\&
+ L^{(2,3)}_n \lambda_n ((-1 +
            r_n) (\alpha^R_n + \beta^R_n) + \beta^I_n (-1 + \alpha^R_n + \
\beta^R_n) \lambda_n) \nonumber\\&+
      L^{(1,3)}_n (-1 + \alpha^R_n + r_n \alpha^R_n + \beta^R_n +
         r_n \beta^R_n + \beta^I_n \lambda_n - (1 +
            2 r_n) \beta^I_n (\alpha^R_n + \beta^R_n) \lambda_n))\Big]\Big/
\nonumber            \\&
            2 \big(L^{(1,1)}_n
- L^{(1,3)}_n r_n + L^{(3,3)}_n r_n^2 + \lambda_n (L^{(1,2)}_n - L^{(2,3)}_n r_n + L^{(2,2)}_n \lambda_n)\big)\nonumber,\\
L_{n-1}^{(3,3)}\nonumber &=
-\Big[(L^{(1,3)}_n)^2 (-1 + \alpha^R_n + r_n \alpha^R_n + \beta^R_n + r_n \beta^R_n)^2 +
     2 L^{(1,3)}_n \lambda_n \big((-1 + \alpha^R_n + r_n \alpha^R_n + \beta^R_n +
           r_n \beta^R_n)\times \\&\big(L^{(2,3)}_n (-1 + \alpha^R_n + \beta^R_n) \nonumber-
           L^{(1,2)}_n (\alpha^R_n + \beta^R_n)\big) -
        2 L^{(2,2)}_n (-1 + \alpha^R_n + \beta^R_n) (\alpha^R_n + \beta^R_n) \
\lambda_n\big)
\nonumber\\&-
     4 L^{(1,1)}_n \Big(L^{(3,3)}_n (-1 + \alpha^R_n + r_n \alpha^R_n + \beta^R_n +
           r_n \beta^R_n)^2 + (\alpha^R_n + \beta^R_n) \lambda_n \times\nonumber\\&\big(-L^{(2,3)}_n (-1 \
+ \alpha^R_n + r_n \alpha^R_n + \beta^R_n + r_n \beta^R_n) +
           L^{(2,2)}_n (\alpha^R_n + \beta^R_n) \lambda_n\big)\Big)
\label{L33}          \\&
           + \lambda_n \Big(\big((L^{(2,3)}_n)^2 \
- 4 L^{(2,2)}_n L^{(3,3)}_n\big) (-1 + \alpha^R_n + \beta^R_n)^2 \lambda_n +
        (L^{(1,2)}_n)^2 (\alpha^R_n + \beta^R_n)^2 \lambda_n\nonumber \\&-
        2 L^{(1,2)}_n (-1 + \alpha^R_n + \beta^R_n) (2 L^{(3,3)}_n (-1 + \alpha^R_n +
              r_n \alpha^R_n + \beta^R_n + r_n \beta^R_n) -
           L^{(2,3)}_n (\alpha^R_n + \beta^R_n) \lambda_n)\Big)\Big]\Big/
\nonumber           \\
           &
           4 \big(L^{(1,1)}_n - L^{(1,3)}_n r_n +
       L^{(3,3)}_n r_n^2 + \lambda_n (L^{(1,2)}_n - L^{(2,3)}_n r_n + L^{(2,2)}_n \lambda_n)\big).\nonumber
\end{align}

\end{footnotesize}

\end{document}